\documentclass[12pt]{article}
\usepackage{graphicx}

\newcommand{\ds}{\displaystyle}
\newcommand{\be}{\begin{equation}}
\newcommand{\ee}{\end{equation}}
\newcommand{\bea}{\begin{eqnarray} \displaystyle}
\newcommand{\eea}{\end{eqnarray}}

\newcommand{\ba}{\begin{array}}
\newcommand{\ea}{\end{array}}
\newcommand{\besys}[1]{\begin{equation} #1 \left\{ \begin{array}{l} \displaystyle}
\newcommand{\eesys}[1]{\end{array} \right. #1 \end{equation}}

\newcommand{\inv}[1]{\frac{1}{#1}}
\newcommand{\const}[3]{\left(\frac{#1}{#2}\right)^{#3}}

\newcommand{\prt}[1]{\left(#1\right)}

\newcommand{\lsim}{\ensuremath{\raisebox{-1.5 mm}{$~\stackrel{\ds <}{\sim}~$}}}
\newcommand{\gsim}{\ensuremath{\raisebox{-1.5 mm}{$~\stackrel{\ds >}{\sim}~$}}}
\newcommand{\de}{\partial}

\newcommand{\mach}{\ensuremath{\cal M}}

\newcommand{\eps}{\ensuremath{\epsilon}}

\newcommand{\bnab}{\mbox{\boldmath $\nabla$}}

\newcommand{\bu}{\mbox{\boldmath $u$}}
\newcommand{\bx}{\mbox{\boldmath $x$}}

\newcommand{\yr}{{\rm yr}}
\newcommand{\msolar}{\ensuremath{{\rm M}_\odot}}
\newcommand{\lsolar}{\ensuremath{{\rm L}_\odot}}

\newcommand{\X}{\ensuremath{\underline{X}}}

\renewcommand{\mach}{{\rm Ma}}

\begin{document}

\title
{Gas Dynamic Stripping and X-Ray Emission of Cluster Elliptical Galaxies.}
\author{Thomas Toniazzo,\\
Department of Physics \& Astronomy, University of Leeds, UK 
\and
\&\\
Sabine Schindler, \\
Astrophysics Research Institute, Liverpool John Moores University,\\
Twelve Quays House, Birkenhead CH41 1LD, UK}
\date{Draft January 2001}

\maketitle

\abstract{
Detailed 3-D numerical simulations of an elliptical galaxy orbiting in a 
gas-rich cluster of galaxies indicate that gas dynamic stripping is less 
efficient than the results from previous, simpler calculations (Takeda et 
al. 1984; Gaetz et al. 1987) implied. This result is consistent with X-ray 
data for cluster elliptical galaxies. 
Hydrodynamic torques and direct 
accretion of orbital angular momentum can result in the formation of 
a cold gaseous disk, even in a non-rotating galaxy. 
The gas lost by cluster 
galaxies via the process of gas dynamic stripping 
tends to produce a 
colder, chemically enriched cluster gas core. A comparison of the models 
with the available X-ray data of cluster galaxies shows that the X-ray 
luminosity distribution of cluster galaxies may reflect hydrodynamic 
stripping, but also that a purely hydrodynamic treatment is inadequate 
for the cooler interstellar medium near the centre of the galaxy. }

\section{Introduction}

Normal elliptical galaxies contain hot coronal gas which emits 
radiation in 
the X-ray band through thermal Bremsstrahlung and collisional line excitation. 
Total gas masses of the hot interstellar medium (ISM) 
in elliptical galaxies vary between $10^8$ and 
$10^{10}\msolar$, and typical gas temperatures are $T\lsim 1$ keV, 
implying that radiative cooling is efficient.

Although, in general, larger elliptical galaxies are also 
brighter in X-rays, the hot-gas content of elliptical galaxies proves to be 
very variable, as can be inferred from a scatter plot of the X-ray luminosity
$L_X$ (in a given waveband) and the blue optical luminosity $L_B$ for the 
observed galaxies (e.g. Irwin \& Sarazin 1998). 
This fact can be related either to the 
persistence, or otherwise, of a galactic wind sweeping the gas shed from the 
stellar population of the galaxy (Ciotti et al. 1991), or to the 
different environments where the galaxies reside. 
The apparent lack of  
correlation between the residuals of the $L_B$--$L_X$ relation and any 
intrinsic properties of the galaxies in other wavebands may 
favour the 
second explanation (White \& Sarazin 1991). A gas-rich cluster environment can 
cause gas dynamic gas-mass loss from (or ``stripping'' of) the galaxy 
(Schipper 1974), 
or also promote radiative 
cooling by pressure confinement and eventually produce an accretion flow 
onto the galaxy (Ciotti et al. 1991). The effect of the environment therefore 
is likely to depend on the size and the depth of the potential well of
the galaxy  
(Bertin \& Toniazzo 1995) and on its motion relative to the intra-cluster 
medium (ICM). 

Gas dynamic stripping 
may be related to the chemical enrichment of the ICM, and possibly also its 
dynamics (Deiss \& Just 1996). 
A
preliminary conclusion 
from the analysis of X-ray spectra of the ICM is 
that the metal abundances in the ICM 
vary from one type of cluster to another, and that the distribution of 
metals within clusters, especially ``cooling flow'' clusters, 
can be non-uniform (Finoguenov, David \& Ponman 1999; Irwin \& Bregman
2000; De Grandi \& Molendi 2000). 
This 
suggests that both winds 
(mainly responsible for [Si] enrichment) 
and galaxy stripping 
(resulting mainly in [Fe] enrichment)
contribute to the chemical 
enrichment of the ICM to variable extents, possibly according to the 
structure of the cluster and the age of the brightest cluster galaxies. 

Previous theoretical investigations 
of gas dynamic stripping
(Lea \& De Young 1976; Takeda, Nulsen \& Fabian 1984; Gaetz, Salpeter 
\& Shaviv 1987; Portnoy, Pistinner \& Shaviv 1993; Balsara, Livio, \& 
O'Dea 1994) generally found 
that ram-pressure stripping of cluster galaxies should be 
very effective 
and that a generic cluster elliptical galaxy should retain very little of 
the gas shed by its stars. 
%
Cluster galaxies should thus rarely be very luminous (if detectable)
in X-rays. These conclusions are likely to be affected by some
assumptions which were made to simplify the mathematical or
computational problem. They are not supported by more recent 
X-ray data, which show that, by contrast, most X-ray bright
ellipticals are found within clusters or groups. 
Careful analysis of high-resolution X-ray data reveals more and more X-ray 
bright galaxies against the high background of the cluster-gas emission 
(e.g. Drake et al. 2000). 
%
Such evidence indicates 
that, in general, dense environments 
host {\it both} ISM-depleted and very ISM-rich ellipticals. In the latter, 
the cooling times of the coronal ISM are short. 

Generally speaking, clusters of galaxies at the present 
epoch appear to be still in the process of accreting smaller 
groups of galaxies (Dressler \& Shectman 1988; Neumann 1997)
and single galaxies.
In the present work we ask what the 
hydrodynamic evolution of a galaxy's ISM will be when an elliptical galaxy is
falling into a cluster
of galaxies and is moving with respect to the ICM.

To answer this question and update our theoretical understanding of the ISM 
dynamics of cluster galaxies, a time dependent, three-dimensional, treatment 
of the hydrodynamic problem, including careful modeling of the ``background'' 
galaxy and cluster of galaxies, is desirable. 

The 
generic orbit for a galaxy in a cluster is neither circular nor radial. 
Instead, it is highly elongated, and 
%
%
describes
a rosette-like shape 
(Binney \& Tremaine 1987, \S3.1, p.106). 
As a consequence, the galaxy `visits' regions of the cluster with 
an ICM at different densities, while its velocity changes both in magnitude 
and in direction. Furthermore, given a typical temperature of 5-6 keV for the 
ICM, and a typical velocity dispersion for the cluster of 1000 km/s, it 
follows that a typical galaxy moves with supersonic speed on a part of 
its orbit close to the pericentre, and with a subsonic speed near the 
apocentre. As we show in the next section, this suggests that the dynamics 
of the ISM-ICM interaction is not two-dimensional, that it is not governed 
by ram pressure alone, and that it is not steady. 

While the central cooling time scale $\tau_c$ in 
an average 
elliptical galaxy is much longer than the 
dynamic scale for transonic stripping, 
optically thin cooling at $T\sim1$ keV has the tendency of proceeding in a 
run-away fashion in which $\tau_c$ decreases even further unless the 
gas is {\it completely} removed or reheated. When supersonic stripping pauses, 
the amount and concentration of gas in the galaxy is crucial in determining 
its further evolution. 
This fact, together with the 
greater complexity of a three-dimensional flow, makes simplistic 
analytical estimates of the efficiency 
of ``ram-pressure'' stripping of spiral galaxies 
(Gunn \& Gott 1972; Abadi et al. 1999) 
unreliable when applied to elliptical galaxies.

Due to the strong feedback of cooling on the dynamics of stripping, 
one should be careful with allowing for gas
``drop out'' -- gas removal due to radiative cooling -- 
as did Gaetz et al. (1987), Portnoy et al. (1993) and Balsara et al. (1994). 
Not only does it affect the gas dynamics (Portnoy et al. 1993) 
but, as we will show, also the long-term evolution of the ISM in the 
time-dependent problem. On one hand, as long as there are no 
observational detections of star-formation rates in cluster ellipticals, 
and there is no compelling theoretical reason for drop out to occur, its 
inclusion in a time-dependent hydrodynamic calculation is somewhat 
arbitrary. On the other hand, if stripping was efficient {\it regardless} 
of drop out, neither star formation nor drop out would be needed anymore in 
the present context. 

We perform the first three-dimensional, time-dependent hydrodynamic
numerical calculations of this problem. We start from
a prescribed initial state (similar to that of Lea \& De Young 1976)
and include the full radiative cooling due to continuum and line emission.
We generally do not allow for drop out, 
except for comparison with common 
hydrodynamic models of the X-ray gas (e.g. Sarazin \& Ashe 1989). 
The model we adopt to represent the galaxy (light distribution, 
gravitational potential) undergoing the stripping process is roughly 
consistent 
with well-constrained stellar dynamic models of known cluster elliptical 
galaxies (Saglia, Bertin \& Stiavelli 1992; Dehnen \& Gerhard 1994). 
Mass and energy sources are treated in a standard 
fluid-average way valid over sufficiently large times and lengths. 
The cluster ICM and gravitational potential are chosen to be 
similar to the properties of the Coma cluster. 
As a result of such detailed modeling, our computations depend on a 
considerable 
number of assumptions and parameters, which, however, are not fine tuned 
but matched as far as possible to current observational knowledge. 
In order to check the validity of the assumptions and to 
estimate the reliability of the results in terms of their astronomical 
implications, we explicitly compare the model results to published X-ray data
on bright cluster ellipticals. 

The paper is organized as follows. 
In Section 2 we discuss the time scales of the problem
and the main hydrodynamic 
mechanisms at work in the supersonic flow and in the subsonic flow. 
In Section 3 we present the relevant equations (Sect. 3.1), the underlying 
models necessary to calculate the source terms and the gravitational field 
(Sect. 3.2), and the method of their solution (Sect. 3.3). 
We discuss also (Sect. 3.3.2) the initial condition and (Sect. 3.3.3) 
the boundary conditions adopted. 
In Table 1 the parameters adopted for the four models are summarized. 
Section 4 contains the main results for the hydrodynamic evolution 
of the ISM in the four models. In Sections 4.1 and 4.2 we focus on the 
supersonic and the subsonic part, respectively, in Section 4.3 we discuss the 
results when drop out is included, and Section 4.4 concerns 
the long-term evolution determined by the combined effect of 
the progressive depletion of the outer ISM halo and of the cooling 
taking place at the galaxy centre. 
In Section 5 the consequences of stripping on the thermal evolution of 
the ISM are discussed, with emphasis on the resulting X-ray properties 
of the halo. In Section 6 we present the computed X-ray morphologies and 
compare them to X-ray data of well-studied cluster galaxies. 
In Section 7 we derive some possible implications of the stripping 
process of cluster galaxies on the ICM. Finally, Section 8 summarizes.

\section{Time scales and dimensionality of the hydrodynamic problem}

The time evolution of dynamic stripping is governed by five characteristic 
time scales, which are given by the crossing time 
$\ds \tau_{cr}=\frac{2L_{gr}}{c_h\mach}$, the free-fall time 
$\ds \tau_{ff}=\frac{ L_{gr}}{\sqrt{3}\sigma_*}$, the cooling time
$\ds \tau_c   =\frac{ 3 n T }{2n_in_e\Lambda(T)}$, the
Kelvin-Helmholtz (K-H) linear 
growth time $\ds \tau_{KH}\simeq\frac{3L_{gr}}{c_h}$, and the replenishment time
$\ds \tau_{rep}=\frac{\rho}{\dot\rho_{inj}}$. We have indicated with $L_{gr}$ 
the gravitational length-scale 
$L_{gr}\equiv GM/c_s^2$, where $M$ is the total 
gravitational mass of the galaxy and $c_s$ is the sound speed in the 
ISM. The quantity 
$\mach$ is the Mach number, $U_G/c_h$, of the galaxy in the ICM (the more 
conventional symbol $M$ being used to 
indicate a mass); $c_h$ is the sound speed 
in the ICM, and typically $c_s<c_h$; $\dot\rho_{inj}$ is the 
mass-injection rate due to stellar deaths; and $\sigma_*$ is the average 
value of the one-dimensional velocity dispersion of the stars in the galaxy 
which characterizes
the free-fall velocity in the gravitational potential of the galaxy. 
The factor 3 
in the expression for $\tau_{KH}$ is approximate since the K-H growth rate is 
approximately proportional to $\mach$ 
for $\mach>1$ 
and to $c_h/c_s$ for $\mach<1$ 
and $c_h/c_s<1$. Finally, the expression $n_in_e\Lambda$ in the cooling time 
represents the radiated power per unit volume 
(Eq. (\ref{coolfact}) in  section 
\ref{govern}). Numerically, these time scales are
\bea
 \tau_{cr} = & \ds \frac{2L_{gr}}{c_h\mach} & = 47~  {\rm Myr~} \, \mach^{-1}
                   \const{M}{10^{12}\msolar}{} \const{T_s}{\rm keV}{-1}
                   \const{T_h}{\rm 8 keV}{-1/2} \, ,
\\ \nonumber \\
 \tau_{KH} = & \ds \frac{3L_{gr}}{c_h} & = 70~  {\rm Myr~} \,
                    \const{M}{10^{12}\msolar}{} \const{T_s}{\rm keV}{-1}
                    \const{T_h}{\rm 8 keV}{-1/2} \, ,
\\ \nonumber \\
 \tau_{ff} = & \ds \frac{L_{gr}}{\sqrt{3}\sigma_*} & = 49~  {\rm Myr~} \,
                   \const{M}{10^{12}\msolar}{} \const{T_s}{\rm keV}{-1}
                   \const{\sigma_*}{\rm 300 km/s}{-1} \, ,
\\ \nonumber \\
 \tau_c    = & \ds \frac{3}{2}\frac{n}{n_en_i}\frac{T}{\Lambda(T)} 
              & = 540~ {\rm Myr~} \, \const{T}{\rm keV}{1.65} 
                    \const{n}{10^{-2} {\rm cm}^{-3}}{-1} \, ,
\\ \nonumber \\
 \tau_{rep} = &\ds \frac{\rho}{\dot\rho_{inj}} & = 430~ {\rm Myr~} \, 
                    \const{   n  }{ 10^{-2}   {\rm cm}^{-3}}{}
                    \const{\rho_*(r_*)}{10^{-24} {\rm g~cm}^{-3}}{-1}
                    \const{r}{r_*}{p}  \, .
\eea
Here, $T_s$ and $T_h$ are the temperatures of the ISM and the ICM, respectively; 
$r_*$ is a scale radius for the galactic stellar matter distribution, and $p$ is 
an exponent which varies between $p=4$ for $r>r_*$ and $p=2$ for $r<r_*$ (Jaffe 
1983). 
Three of the given time scales are `fast', related to the dynamics of pure 
ablation, pure shear and pure infall, respectively, while the remaining two are 
`slow', related to local cooling and local mass injection. Because of its 
dependence on position and on the gas density, mass injection is dynamically 
relevant near the centre of the galaxy also when stripping is efficient, 
contributing to 
the formation of a bow shock in front of the galaxy. 
In the time-dependent stripping problem, cooling, 
which is associated with a long time scale, 
is unimportant only if stripping is complete. 
In a supersonic flow the gas can be displaced and eventually removed from the 
galaxy if the condition 
\be \rho_h c_h^2 \mach^2 \ge \int\rho_s\bnab\Psi_G\cdot d{\bf l} 
\label{spannometrica}\ee
is satisfied (Takeda et al. 1984) for a time of the order of a few crossing times
$\tau_{cr}$. In Eq. (6), $\Psi_G$ is the galactic gravitational potential, 
the integral being taken along a path in the direction of the freestream velocity. 
If the galaxy is not stripped completely, during the time between one pericentre 
passage and the next, cooling may cause $\rho_s$ to increase enough
close to the 
centre (where the gravitational force is strongest) to invalidate Eq. 
(\ref{spannometrica}). 
The competition between gas cooling, accretion and gas removal 
in the subsonic part of the orbit is therefore crucial in determining the evolution 
of the ISM. The denser ICM near the centre of the 
cluster causes both ablation of the outer layers and compression of the ISM in 
the centre, thus favouring central cooling. If cooling proceeds
quickly enough, 
further evolution will be shifted by the very short central cooling time 
towards net accretion of gas. 
In general, a given hydrodynamic regime 
can last only over a time comparable to the time over which the kinetic pressure 
acting on the ISM does not change significantly, i.e. 
\bea
\tau_{rp} & = \ds \left[\frac{d~}{dt}\ln(\rho_CU_G^2)\right]^{-1} \nonumber \\
       &\simeq \ds \frac{r U_G}{\sigma_C^2}\left(2+\mach_r\right)^{-1}   \nonumber \\ & 
        \simeq 130~ {\rm Myr~} \, \const{r_C}{500{\rm~kpc}}{}
             \const{T_h}{8{\rm~keV}}{1/2}\const{\sigma_C}{1200{\rm~km/s}}{-2}
\, .\eea
This time scale is neither very long compared to the `short' time scales, nor 
very short compared to the `long' time scales, and implies that  
the resulting hydrodynamic problem is non-steady and non-linear. 

For a general orbit of the galaxy, the time scale $\tau_{rp}$ also
determines the 
rate at which the direction of the free-stream speed varies, modifying 
the force equilibrium in the ISM. 
Since $\tau_{rp}$ is shorter than the cooling time, the distribution of the 
ISM in the galaxy is also affected by the two-dimensional orbital motion of 
the galaxy. 
Finally, when the Reynolds number is not small, subsonic stripping depends 
on the development of Kelvin-Helmholtz modes and on eddy formation, which 
produce a non-axisymmetric flow even around a galaxy in steady motion 
along its symmetry axis. The generation of vortices is important in 
determining the structure of the wake left by the galaxy. 
A three-dimensional treatment is therefore necessary in order to
estimate correctly 
the drag force on the ISM and the mixing between ISM and ICM. 

\section{Description of the Model}


\subsection{Governing Equations\label{govern}}

The hot interstellar medium is treated as a Eulerian (non-viscid) ideal 
monoatomic gas with mass density $\rho$, pressure $p$, temperature $T$ 
(in energy units) and velocity $\bu$. 
The time evolution of these quantities is followed in the reference system 
of the galaxy which is subject to the acceleration ${\bf g}_G$ with respect to 
the inertial frame of the cluster. The set of equations which
determines the time  
evolution is defined by the hydrodynamic equations:
\bea
\de_t\rho + \bnab\cdot\rho\bu  &=&   \dot\rho_{inj}  \, , 
\label{mass} \\ &&\nonumber \\ \ds
\de_t\bu +\prt{\bu\cdot\bnab}\bu + \inv{\rho}\bnab p  &=& - \bnab\Psi_G 
     + {\bf g}_G-\bnab\Psi_C
     + \frac{\dot\rho_{inj}}{\rho}\prt{\bu_{inj}-\bu}    \, ,
\label{euler} \eea
\bea
\de_t\left[\rho\left(\frac{1}{2}u^2+\frac{3T}{2\mu}+\Psi_G\right)\right] + 
\bnab\cdot\left[\rho\bu\left(\frac{1}{2}u^2+\frac{3T}{2\mu}+\frac{p}{\rho}+
   \Psi_G\right)\right] = \mbox{\hspace{2cm}} && \nonumber \\ \ds
\mbox{\hspace{2cm}}
  \dot\rho_{inj}\left(\frac{1}{2}u_{inj}^2+\Psi_G\right) 
+ \rho\bu\cdot({\bf g}_G-\bnab\Psi_C) 
+ \dot\eps_{inj}-\dot\eps_{rad}  \, ,
\label{enthalpy} \eea
complemented by the equation of state 
\be
 p \,=\, nT \label{eos} \, , \ee
and the relations
\bea
\mu &=& \rho/n \label{defmu} \, , 
\\ \ds
 n  &=& n_i+n_e \label{sum} \, ,
\\ \ds
n_i &=& \frac{\rho}{m_p}    \sum_Z\frac{ X_Z}{A_Z} \; , 
\\ \ds
n_e &=& \frac{\rho}{m_p} f_e\sum_Z\frac{ZX_Z}{A_Z} \, .  \eea
The symbols $\Psi_G$ and $\Psi_C$ represent the gravitational potential 
fields generated by the mass distribution of the galaxy and by the mass 
distribution of the cluster, respectively. We neglect the self-gravity 
of the gas, so that both $\Psi_G$ and $\Psi_C$ are given functions of 
position determined by the assumed galaxy and cluster models. 
The combination ${\bf g}_G-\bnab\Psi_C$ is a tidal field, with 
the fictitious acceleration ${\bf g}_G$ given by 
\be {\bf g}_G = - \bnab_{{\bf R}_G}\Psi_C({\bf R}_G) \, , \label{fict}\ee
where ${\bf R}_G=-{\bf x}_C$ is the position vector of the galaxy with respect to 
the centre of the cluster. 

The quantities $n$, $\mu$, $n_i$, $n_e$, $m_p$, $f_e$, $A_Z$ and $X_Z$ 
indicate the total particle number density, the mean mass per particle, 
the total ion number density, the electron number density, the average 
mass of one nucleon $m_p=1.67\times10^{-24}$ g, the electron number fraction 
$n_e/n_i$, the atomic weight of atoms with atomic number 
$Z$, and the weight fraction of the atomic species with the atomic number 
$Z$. The mass of each species is varied only through the distributed 
mass sources, so that each of the quantities $X_Z$ obeys the evolution 
equation
\be
\de_t\prt{\rho X_Z} + \bnab\cdot\prt{\rho X_Z\bu} = 
                             \dot\rho_{inj} X_{Z,inj}  \, .
\label{tracers}\ee
In the present work only 
one
of the $X_Z$'s 
is 
varied independently, 
namely Fe (N=26). 
The other $X_Z$'s are taken to reflect a 
mixture of ``solar'' (Anders \& Grevesse 1989) and ``cosmic'' (Wheeler et al. 
1989) abundances determined by the ratio [O]/[Fe]. For this reason in our 
models [Si]/[Fe] = 0.694 [O]/[Fe] + 0.305 in solar units. 

The optically thin, collisional radiative cooling term is expressed as 
usual in terms of a cooling function $\Lambda$: 
\be \dot\eps_{rad} = n_e n_i \Lambda(T,\X) \, , \label{coolfact} \ee
where $\X=\{X_{\rm O},X_{\rm Fe}\}$. 
We adopt the cooling functions given by Sutherland \& Dopita (1993) 
for conditions of non-equilibrium ionization. For gas cooling at $T<0.1$ keV, 
ionization equilibrium is not achieved and gas cooling rates are smaller than 
those derived under that assumption. At higher temperatures, the equilibrium 
and the non-equilibrium cooling functions are essentially indistinguishable. 
The cooling coefficient for given $(T,\X)$ is calculated 
using log-linear interpolation in the temperature $T$, and linear interpolation 
in the abundance levels. The latter is a fairly good approximation for sub-solar 
abundances. For $T>$2.7 keV, we extrapolate 
the cooling function using the pure Bremsstrahlung law $\Lambda\sim T^{1/2}$. 

Usually, in steady-state models for the X-ray gas (including the 
simulation work on steady-state stripping) a sink term 
$-\dot\rho_{do}$ representing gas ``drop out'' is added on the 
right-hand side of Eq. (\ref{mass}), with corresponding terms in the 
energy equation under the assumption that there is no heat exchange between 
the drop outs and the gas in the flow. Defining the local gas cooling time 
$\tau_c=(3nT/2)/\dot\epsilon_{rad}$, we adopt a prescription of the form 
\be \dot\rho_{do} = q\rho/\tau_c \, , \label{dropout}\ee
which is found to emulate reasonably well the dynamics of the ISM in X-ray 
bright galaxies (Sarazin \& Ashe 1989; Bertin \& Toniazzo 1995). However, 
we take $q$ to be different from zero for only one model (``Bq'',
with $q=0.4$). 


Finally, the source terms are given functions of the coordinates: 
\be
 \dot\rho_{inj} = \alpha_*\rho_*(\bx) 
\, , \label{rhoinj}\ee
\be 
 \dot\eps_{inj} = \alpha_*\rho_*(\bx)
        \left[\sigma_R^2(\bx)+\sigma_\phi^2(\bx)/2+\sigma_{SN}^2\right]
\, . \label{epsinj}\ee 
Here, $\alpha_*$ is the rate of mass return from a passively aging 
stellar population (Renzini 1988), and it is given by 
\be
\alpha_*= 4.75\times10^{-19}\left(\frac{M_*/L_B}{\msolar/\lsolar}\right)^{-1}
      \left(\frac{\tau_*}{15{\rm Gyr}}\right)^{-1.3}
~{\rm s^{-1}}\, , \label{Renzini}\ee 
as a function of the presumed thermonuclear age $\tau_*$ of the system, 
and of the present day (i.e. $\tau_*$=15 Gyr, by definition) 
mass-to-luminosity ratio of the stellar population. We take 
$\tau_*=15$ Gyr initially  for all models. The stellar mass density, $\rho_*$, 
and the one-dimensional stellar velocity dispersion components of the stars, 
$\sigma_R$ and $\sigma_\phi$, are derived from the models described in the next 
Section. The additional heating term $\propto\sigma_{SN}^2$ is due to Supernovae 
of type Ia which are supposed to be distributed according to the stellar matter. 
We assume that each SNIa deposits into the ambient gas an energy of 
$8\times10^{50}$ erg. In terms of the rate of SNIa events (events per century per 
$10^{10}\lsolar$ blue luminosity), 
$\ds \sigma_{SN}=1580\,(r_{SN})^{1/2}\,(4.75\times10^{-19}s^{-1}/\alpha_*)$ km/s. 
We take $r_{SN}=0.06$ for $\tau_*=15$ Gyr and a Hubble constant $H_0=50$km/s/Mpc 
(Cappellaro et al. 1997). We assume that the ratio of SNIa luminosity to stellar 
luminosity does not vary in time, so that $\sigma_{SN}$ is constant. Different, 
and contrasting, assumptions are made in Ciotti et al. (1991) and in Loewenstein 
\& Mathews (1987). The time evolution of the SNIa rate is not known, and our 
approach is to explore the effects of gas dynamic interactions only and neglect 
this possible, additional evolutionary effect (see the discussions in Ciotti et al. 
1991 and Loewenstein \& Mathews 1987). 

Finally, both the initial ISM (see Section 3.3.2) and the newly injected gas 
are assumed to have solar abundances (Anders \& Grevesse 1989), with 
[Fe]/[H]$=2.62\times10^{-3}$ by weight. The ICM, instead, has ``cosmic''
abundance  ratios (Wheeler et al. 1989) and [Fe]$=0.1~$[Fe]$_\odot$ by weight.

\subsection{Input modeling}

\subsubsection{Galaxy model}

The galaxy is represented by the sum of two spheroidal, axisymmetric 
components, of luminous matter and of dark matter respectively, with 
densities
\be \rho_{comp} = \frac{(3-\gamma)M_{comp}r_{comp}}{4\pi d_{comp}} 
 m^{-\gamma_{comp}} (m+r_{comp})^{-4+\gamma_{comp}} 
\label{stellar-distrib}\ee
where $m^2 = R^2 + z^2/d_{comp}^2$ in cylindrical coordinates, $1-d_{comp}^2$ 
is the eccentricity and the subscript $comp$ denotes either the
luminous 
($comp=*$)
or the dark 
($comp=D$)
component. 
We based our treatment on Dehnen \& Gerhard (1994), 
with the inclusion of a second (dark) component, a 
finite inner core radius (i.e. surface of constant $m$), and a finite cut-off 
radius. 
The gravitational potential $\Psi$ and the velocity dispersion profiles $\sigma_R$ 
and $\sigma_\phi$ (in the radial and in the azimuthal direction, respectively) 
are obtained from Eq.(\ref{stellar-distrib}) using Poisson's equation and Jeans' 
equation, respectively (cf. Binney \& Tremaine 1987, \S\S 2.3 and 4.2). 
We assume that both the stellar and the dark matter have zero bulk rotation. 

The values we adopt for the parameters $\gamma_*$, $\gamma_D$, $r_*$, $r_D$, 
$M_*/L_B$ and $M_D/M_*$ 
are taken from the work of Bertin, Saglia \& Stiavelli (1992) and Saglia, 
Bertin \& Stiavelli (1992).
In these models, the dark halo has a mass and an 
extension similar or larger than the matter associated with stars (e.g. Saglia et 
al. 1992; see also Gerhard et al. 1998). 
We take two parameter sets that reflect the properties of 
the elliptical galaxies NGC 4472 (M 49) 
and NGC 4374 (M 84).
These two galaxies are bright in X-rays, with 
typical values of the ratio $L_X/L_B$ for X-ray bright elliptical galaxies. 
The parameters for the M49 model 
are $d_*=d_D=0.8$, $\gamma_*=2$, $\gamma_D=1.1$, $r_*=16.13$kpc, 
$r_D/r_*=2.80$, $M_*=7.12\times10^{11}\msolar$ and $M_D/M_*=1.49$ at a
radius of 150kpc. Those of the M84 model 
are $d_*=d_D=0.9$, $\gamma_*=2$, $\gamma_D=1.85$, $r_*=16.93$kpc, 
$r_D/r_*=1.06$, $M_*=3.49\times10^{11}\msolar$ and $M_D/M_*=1.64$ at a
radius of 100kpc. The blue optical luminosities of the two galaxies 
are $L_B=1.1\times10^{11}\lsolar$ and $L_B=5.8\times10^{10}\lsolar$ for 
NGC 4472 and NGC 4374, respectively. These two models are referred to 
as `GA' and `GB' in Table \ref{partable}. 

\subsubsection{Cluster and Galaxy Orbit\label{cluster}}

We consider a dynamically relaxed, spherically symmetric, rich cluster of 
galaxies (like, e.g. the Coma cluster), and we take the ICM distribution 
to be hydrostatic. The cluster gravitational potential is assumed to be 
\be 
\Psi_C(R) = \sigma_C^2 \ln\left[1+ \left(\frac{R}{r_C}\right)^2\right] 
\, \label{cluster_pot} \ee 
reflecting a mass distribution close to a non-singular, infinite isothermal 
sphere with ``core radius'' for the gravitational mass equal\footnote{This 
is different from the King-type potential 
since the cluster mass density for $R>2r_C$ is flatter than the King-type
distribution ($\rho_{Cluster}\sim1/R^2$ instead of $1/R^3$). We think this 
is appropriate for a cluster which is still accreting objects. It is also 
the reason for the difference by a factor $\sqrt{2/3}$ between the core radius of 
the total gravitational mass and that for the gas distribution.} to $2r_C/3$. 
Here, $R$ is the distance from the centre of the cluster, 
and $\sigma_C$ is the one-dimensional velocity dispersion for the self-gravitating 
mass distribution in the cluster, which gives the velocity scale for the orbital 
motion of the galaxies. Considering the Coma cluster as a reference, we take 
$r_C=500$ kpc (Briel et al. 1992) and $\sigma_C=1201$ km/s (Colless \& Dunn 1996). 

The temperature profile of the ICM is assumed to be parabolic: 
\be T_C = \frac{\sigma_C^2}{\beta_C} 
    \left[1-\xi_C\left(\frac{R}{r_C}\right)^2\right] \, , \label{TC}\ee
and, correspondingly, the ICM density profile is 
\be
\rho_C(R) = \rho_{C0}\, 
    \left[1+\left(\frac{R}{r_C}\right)^2\right]^{-3\bar\beta/2}
    \left[1-\xi_C\left(\frac{R}{r_C}\right)^2\right]^{\bar\beta-1} \, ,
\label{rhoC}\ee
where $\bar\beta\equiv\beta_C/(1-\xi_C)$. For $\xi_C=0$ this would be the 
usual ``$\beta$-profile'' often taken to fit cluster X-ray halos. However, 
the ICM in Coma is known not to be isothermal (Arnaud et al. 2000). For this 
cluster, the central temperature is $\sim 8$ keV while at 800 kpc from the 
centre the temperature can be as low as 5 keV (Arnaud et al. 2000). Since 
8 keV is larger than the ICM temperature of most clusters, we have used two 
different sets of values for $\beta_C$ and $\xi$, one with $\sigma_C^2/\beta_C=6.2$ 
keV (models A1 and A2) and one with higher central temperature and a steeper gradient 
(models B and B$q$ --see Table \ref{partable}).  

The model galaxies describe orbits in the potential $\Psi_C(R)$. The two 
orbits considered are shown in the left panels of Fig. \ref{inflaw}. They 
are defined by two quantities, which can be taken as the pericentre $R_G^{min}$ 
and the apocentre $R_G^{max}$, and an initial position. We always take the latter 
to be the apocentre. The two quantities $R_G^{max}$ and $R_G^{min}$ are given in 
Table \ref{partable} for our models, together with the resulting period of the 
radial motion. 

Note that rich, dynamically relaxed clusters tend to have higher central
densities than those given in Table~1. Such dense systems are usually 
associated to complications such as a central cluster cooling flow, a
large cD galaxy with satellites, and nuclear activity. Our models
therefore do not apply to the central regions of rich clusters. 


\subsection{Method of Solution}

\subsubsection{Numerical scheme}

The problem is represented numerically on a three-dimensional, Cartesian grid of 
points. 


The cell size increases from the centre of the galaxy outwards by a factor 
$p_{m-gr}$ on each successive mesh point. The number of mesh points in each 
spatial dimension, the central cell size $\Delta_0$, and the growth factor 
$p_{m-gr}$ are listed in Table \ref{partable}. For models A2 and B two runs 
were made using different grids, with the aim of comparing the influence of 
the limited resolution on the results.  

The equations are integrated stepwise, alternating conservative hydrodynamic 
Godunov steps and first-order forward time differencing for sources, cooling, and 
cluster tidal forces. The hydrodynamic Godunov code (Godunov et al. 1961) 
used is a version of 
PROMETHEUS (Fryxell, M\"uller \& Arnett 1989) 
which implements the Piecewise Parabolic 
Method (PPM, Colella \& Woodward 1984) and Strang splitting (Le Veque 1998). 
Departures from the equation of state of an ideal gas 
are accounted for in PROMETHEUS 
(following Colella \& Glaz 1985), although for our computations this is only 
a small correction. The gravitational acceleration due to the galaxy is treated 
conservatively (so that the sum of the kinetic energy, the thermal energy and the 
part of the potential energy arising from $\Psi_G$ is conserved by construction), 
while the cluster tidal forces are included as described in Colella \& Woodward 
(1984). Mass and energy sources, cooling and, when present, drop out have been 
included consistently in the general time-centred averaging scheme of PPM. 

\subsubsection{Initial condition}

The computation is started at the orbit's apocentre with an initial gas 
distribution divided in two different regions. For $r<r_{cut}$, the gas is at 
constant temperature $T\sim1$ keV, the metallicity is solar and the velocity (in 
the grid frame, i.e. the galaxy frame) is zero. For $r>r_{cut}$, the temperature 
and density are given by the ICM distribution of the cluster model, as explained 
in section \ref{cluster}, and the metallicity is 1/10 cosmic. 
The galaxy is initially in its apocentre and 
the velocity
of the ICM is the 
opposite of the galaxy's orbital velocity at the apocentre (the
reference system of the calculation is the system of the galaxy).  
The two regions are 
connected smoothly via a spherical shell covering the width of 5 cells. The 
hydrostatic equilibrium condition, i.e. $d(nT)/dr+\rho d\Psi/dr=0$,
holds approximately everywhere. We checked that, if the orbital
velocity was zero, this  
initial configuration was stable. The central temperature and density
of the ISM  
are chosen such that there is approximate equilibrium between heating due to 
energy sources and cooling, and that the central cooling time is comparable 
to the radial orbital period ($\tau_{c,0}=$ ranges from 1.6 to 2.0 Gyr  
while $\tau_{orb}$ ranges from 1.6 to 1.9 Gyr in the different models).

These requirements, which result in a central density 
$n\simeq10^{-2}$ cm$^{-3}$  
and a temperature $T\simeq1$ keV, are motivated by the fact that, on
the one hand, we 
are considering 
a galaxy that has retained a fraction of order 1/2 of the gas shed by its stars 
over its lifetime, and that, on the other hand, when a cooling flow sets in, the 
numerical code loses accuracy and the details of the subsequent evolution become 
unreliable. While any `initial' state is inevitably arbitrary to some extent, 
our choice for the initial halo does have the properties of that of a large, 
gas-rich galaxy which has a central density slightly lower and a central 
temperature slightly higher than those observed in X-ray bright ellipticals. 
Under static conditions, as gas accumulated due to continuous 
injection of matter, the cooling time would become smaller and the heating time 
longer, thus shifting the balance in favour of cooling-driven accretion. 
Basically, we ask of our simulations, what evolution results from the 
competing processes of gas stripping and cooling. 

\subsubsection{Boundary conditions}

The properties of the gas flowing into the grid need to be varied
according to the conditions of the ICM around the galaxy. 
To avoid mismatches at the turning points 
of the motion along, say, the $x$-axis, the boundary conditions at the 
(instantaneous) downstream sides should be kept consistent with those on the 
upstream sides. Thus, we set each quantity at a prescribed value at each 
boundary point, depending on its position on the grid boundary and on the 
instantaneous position and velocity of the galaxy in the cluster as calculated
from its orbital parameters. We made sure that the grid boundaries are
far enough
from the galaxy centre (360 kpc for the full runs) so
that the gas near the boundaries is not affected by the stripping process, 
except within the trail of the galaxy. 
Disturbances artificially generated at the ``end'' 
of the trail could not propagate upstream to the galaxy during 
a half-grid crossing time. 
Over the remaining part of the grid boundary, 
velocity mismatches at the boundaries between the gas within the computational 
grid and that injected into it, caused by the different numerical treatment for 
the acceleration, were minimal and they did not affect the flow near the galaxy. 

The boundary conditions 
set at the grid bottom are less critical, and we choose 
to use the same as those set at the grid sides. For the grid top, we compromised 
between the wish of saving computer time and that of allowing for a true 
three-dimensional flow without imposing symmetry about the equatorial plane which 
leads to a suppression of eddy formation. We therefore placed the upper 
boundary above the symmetry plane of the galaxy 
by at least half of the galactic half-mass radius, 
and imposed planar symmetry beyond that boundary, using non-local reflection 
boundary conditions.

\begin{table}[\protect ht]
\begin{center}
\caption {
Model parameters. 
Cluster parameters: core radius $r_C$, velocity dispersion
$\sigma_C$, central ICM density $\rho_{C0}$, slope of the $\beta$
model $\beta_C$, central ICM temperature $T_{C0}$ and temperature
variation parameter $\xi_C$. Orbit parameters: apocentre $R_G^{max}$,
pericentre $R_G^{min}$ and orbital period $\tau_{orb}$. Galaxy
parameters: model name, scale radius for the galactic stellar matter
distribution $r_*$, mass of the galactic stellar component $M_*$ and blue
optical luminosity $L_B$. Stellar population parameters: thermonuclear
age $\tau_*$ and rate of SNIa events per century per 
$10^{10}\lsolar$ blue luminosity $r_{SN}$. ISM parameters: central ISM
density $\rho_0$, central ISM
temperature $T_0$ and transition radius from ISM to ICM
$r_{cut}$. Mass drop out parameter $q$ according to
Eq. (\ref{dropout}). Grid parameters: number of grid calls in x, y and
z direction $N_x$, $N_y$, and $N_z$, respectively, central cell size
$\Delta_0$, and growth factor $p_{m-gr}$.
\label{partable}}
\hspace{0cm}
\begin{tabular}{l|c|cc|cc}
\hline
\rule[-3mm]{0mm}{3mm}

      & Parameter &    A1  &   $\ds \ba{c}{\rm A2}\\{\rm A2}f\ea$ 
                  & $\ds \ba{c}{\rm B}\\{\rm B}f\ea$ & B{\it q}
\\ \hline
      & $r_C$ (kpc)
                  &   408  &   408  &    408 &    408 
\\ 
      & $\sigma_C$ (km/s)
                  &   1201 &   1201 &   1201 &   1201 
\\ 
      & $\rho_{C0}$ ($10^{-27}$ g/cm$^3$)
                  &   2.13 &   2.13 &  0.165 &  0.165 
\\ 
\raisebox{1.5ex}[-1.5ex]{\bf Cluster}
      & $\beta_C$  
                  &    1.0 &    1.0 &    0.8 &   0.8  
\\ 
      & $T_{C0}$ (keV)  
                  &    6.2 &    6.2 &   7.75 &   7.75 
\\ 
      & $\xi_C$   
                  &  -0.10 &  -0.10 &  -0.13 &  -0.13 
\\ \hline
      & $R_G^{max}$ (kpc)
                  &   1000 &   1000 &   700  &   700  
\\ 
{\bf Orbit}
      & $R_G^{min}$ (kpc)
                  &    360 &    360 &   300  &   300  
\\ 
      &$\tau_{orb}$ (Gyr)
                  &    1.9 &    1.9 &    1.5 &    1.5 
\\ \hline
{\bf Galaxy} 
      & Model name
                  &  `GA'  &  `GB'  &  `GA'  &   `GA'
\\ 
    & $r_*$ (kpc)     &  16.1  &  16.9  &  16.1  &  16.1 
\\
  & $M_*$ ($10^{11}\msolar$) &  7.1   &   3.5  &   7.1  &   7.1
\\
  & $L_B$ ($10^{10}\lsolar$) &  11    &   5.8  &   11   &   11
\\ \hline
{\bf Stellar}
      &$\tau_*$ (Gyr)
                  &     15 &     15 &     15 &     15 
\\ 
{\bf population}
      &$r_{SN}$ (SNU)
                  &   0.06 &   0.06 &   0.06 &   0.06 
\\ \hline
      &$\rho_0$ ($10^{-27}$ g/cm$^3$)
                  &   14.5 &   12.5 &   11.6 &   11.6 
\\ 
{\bf ISM at $t$=0}
      &$T_0$ (keV)
                  &   1.24 &   1.24 &   1.05 &   1.05 
\\ 
      &$r_{cut}$ (kpc)
                  &     65 &     65 &    105 &    105 
\\ \hline 
{\bf Drop-out}
      &$q$
                  &    0.  &    0.  &     0. &    0.4 
\\ \hline
      &$N_x$
                  &    82  &     82 &$\ds\ba{c}144\\100\ea$&     82 
\\ 
      &$N_y$
                  &    82  &     82 &$\ds\ba{c}144\\100\ea$&     82 
\\ 
{\bf Grid}
      &$N_z$
                  &    45  & $\ds\ba{c}45\\71\ea$ & $\ds\ba{c}76\\80\ea$ &  45 
\\ 
      &$\Delta_0$ (kpc)
                  &     2  & $\ds\ba{c}2\\0.5\ea$ & $\ds\ba{c}2\\0.5\ea$ &   2 
\\ 
      &$p_{m-gr}$
                  &  1.063 &  $\ds\ba{c}1.063\\1.04\ea$ &  1.023 &  1.063 
\\ \hline
\end{tabular}
\end{center}
\end{table}


\begin{figure*}[\protect{th!}]
\begin{center}
\begin{tabular}{cc}
\includegraphics[height=6.5cm]{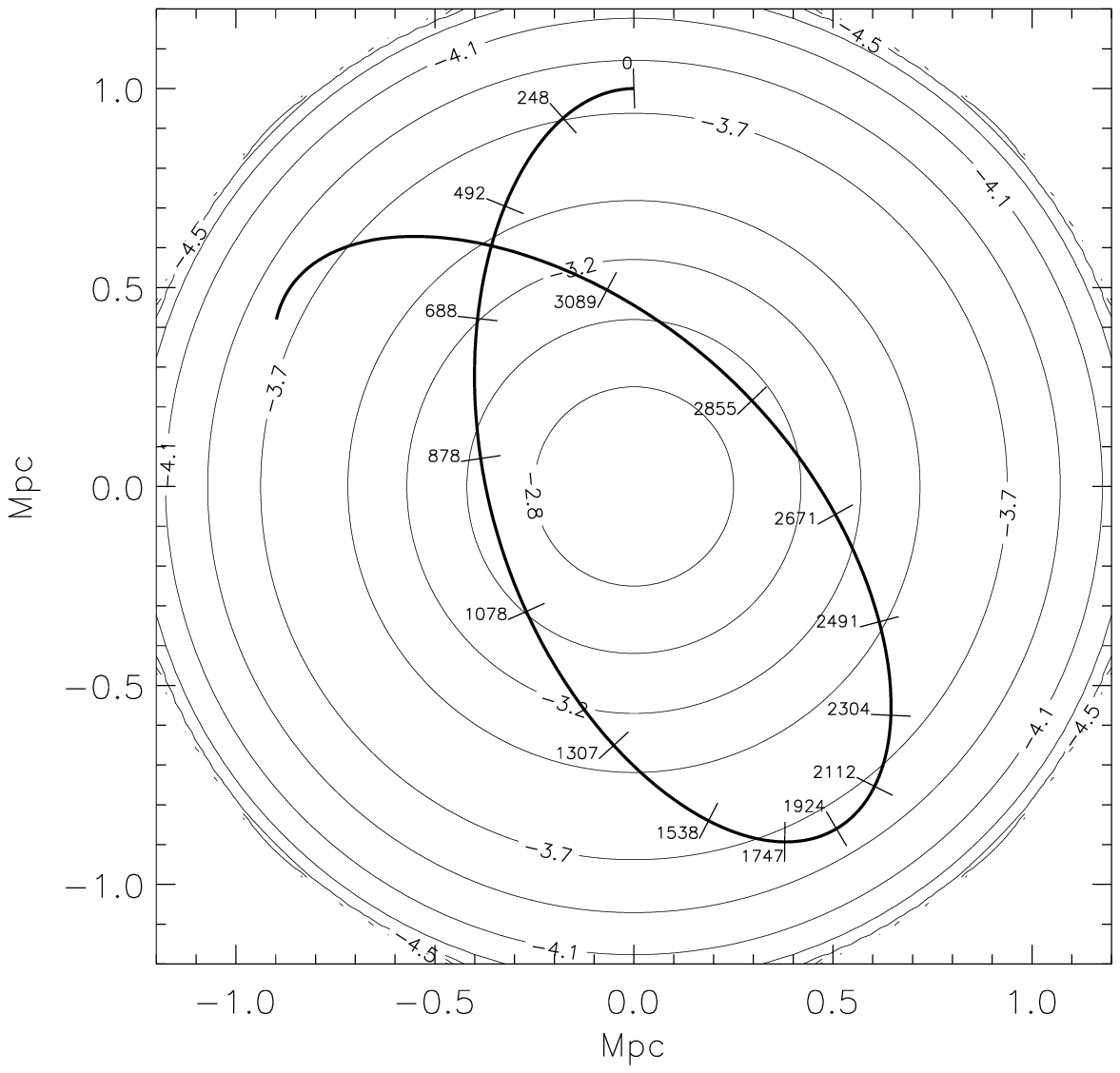}
&
\includegraphics[height=6.5cm]{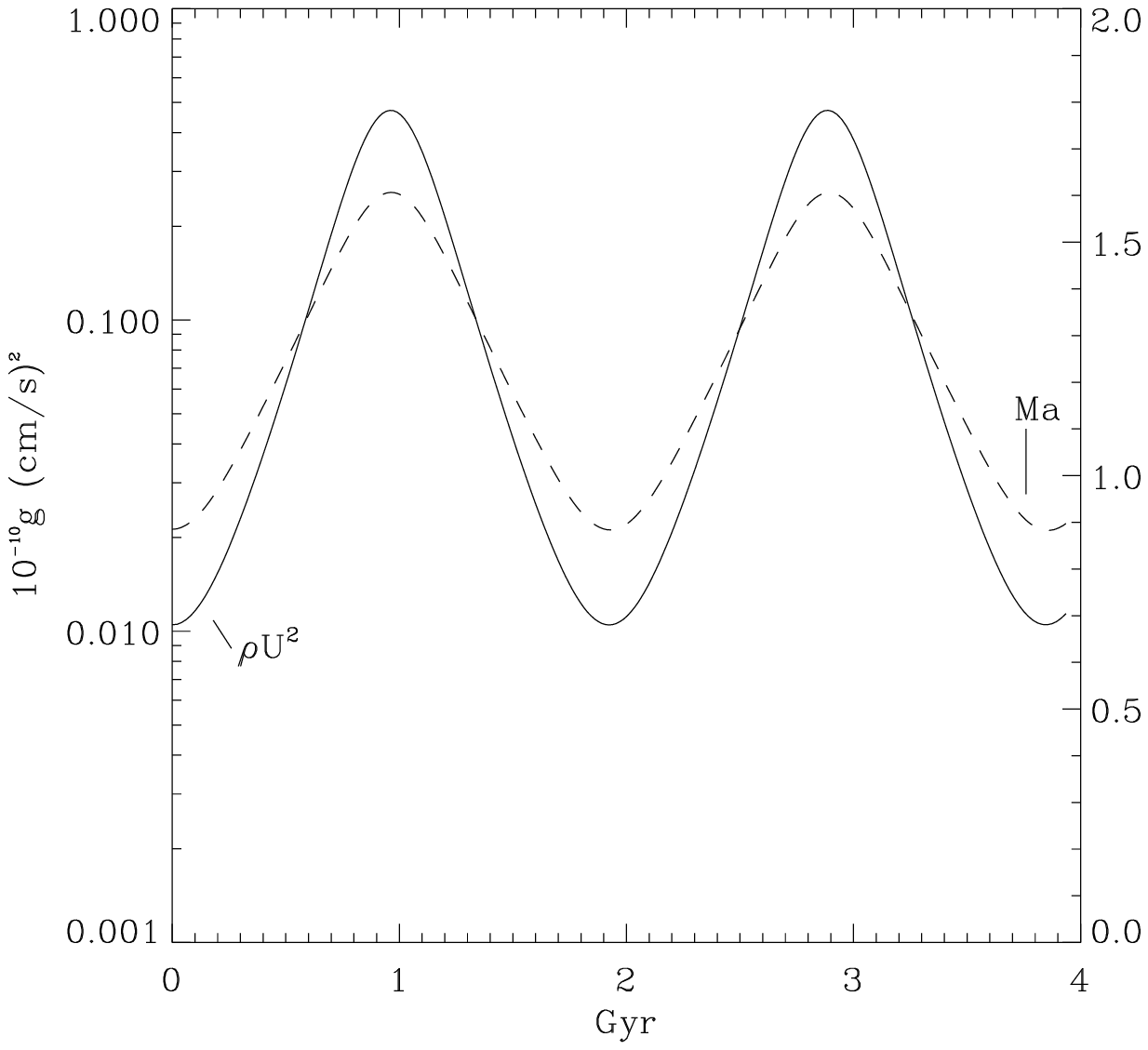}
\\
\includegraphics[height=6.5cm]{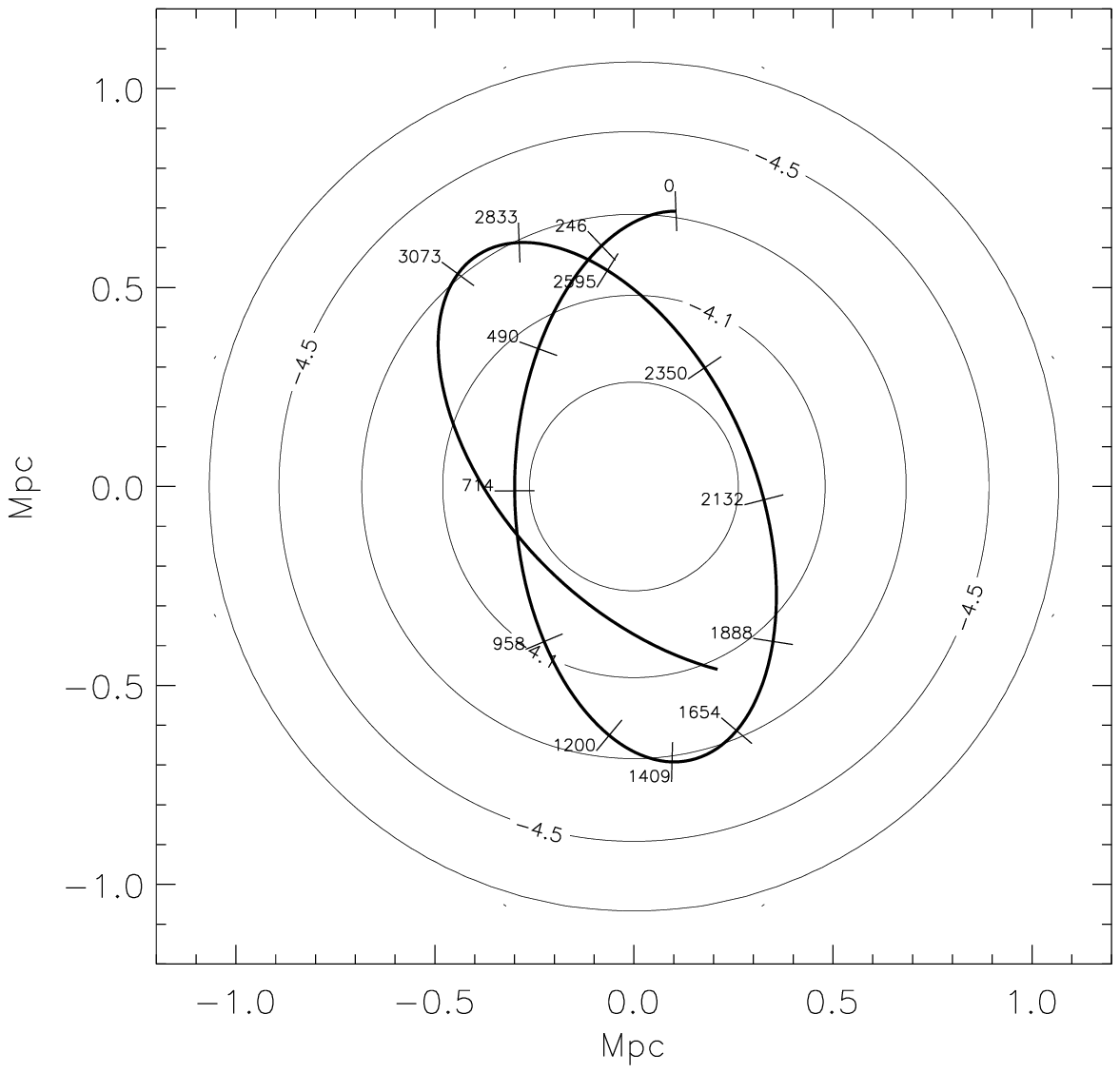}
&
\includegraphics[height=6.5cm]{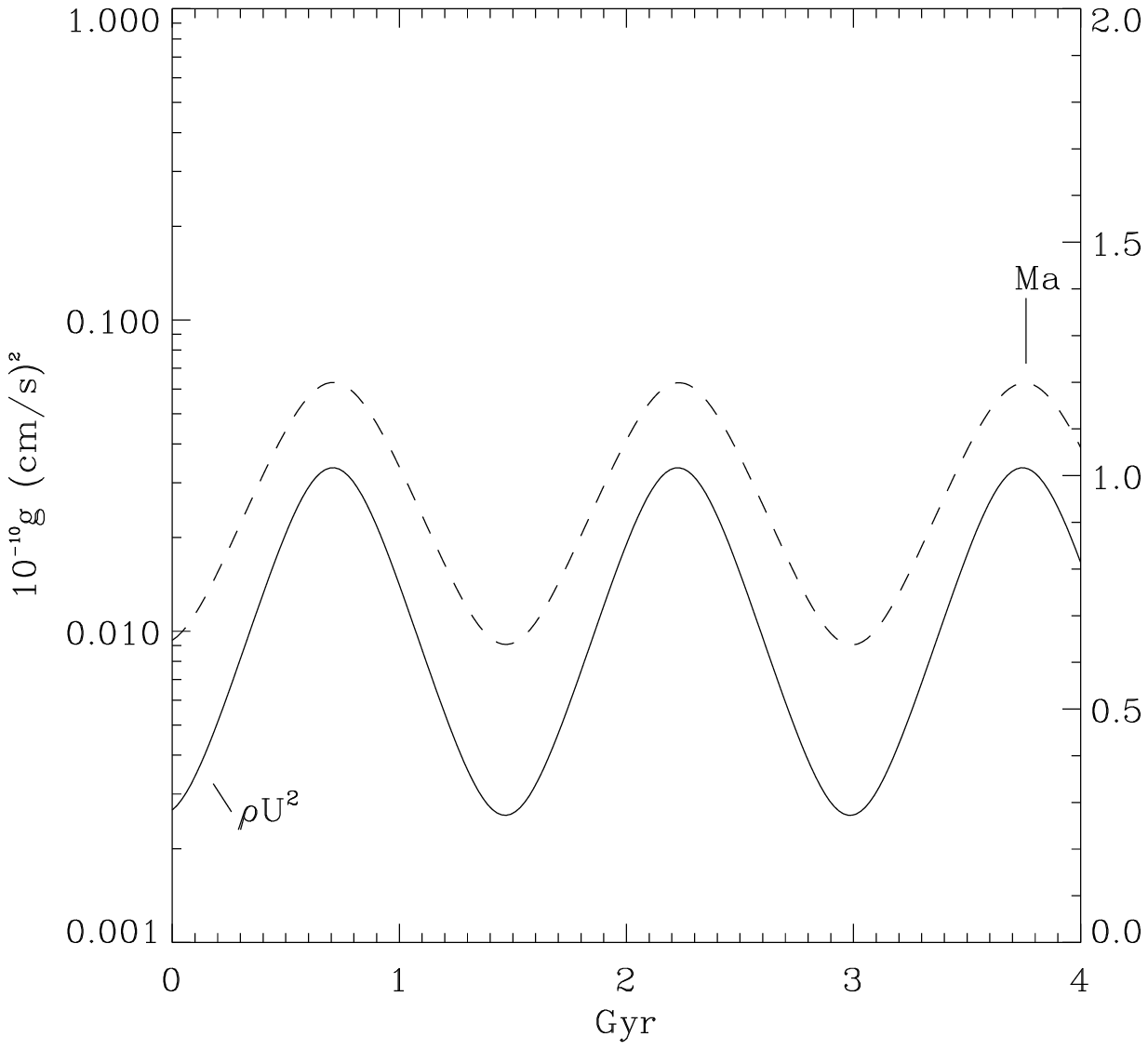}
\end{tabular}
\end{center}
\caption{\small {\it Left:} Galaxy orbital trajectories (thick solid 
lines, labels indicate elapsed time in Myr) in the cluster ICM distribution
(thin solid circles, labels are the decimal logarithm of the ICM density in 
particles/cm$^3$). 
{\it Right:} kinetic force per unit area of streaming ICM in the 
reference system of the galaxy (curves labelled by ``$\rho U^2$'', scale to 
the left) and upstream Mach number (curves labelled ``Ma'', scale on the 
right). The panels in the upper row apply to models A1 and A2, those in 
the lower row to models B and B$q$.\label{inflaw}}
\end{figure*}

\subsection{Model Properties}

The four models presented are intended to enable us to single out 
the factors which are most important 
for the hydrodynamic evolution of the X-ray halo. Relevant model input 
properties, 
in principle, are: 1) the total galaxy luminosity and mass; 2) the density of 
the surrounding gas (ICM) and the velocity (or orbit) of the galaxy; and 
3) drop out. 
Table \ref{partable} lists the model parameters adopted for each of the four 
calculations. Model ``A1'' and model ``A2'' differ only in the galaxy model 
adopted, ``GB'' being about half as luminous and half as massive as ``GA''. 
Model ``A1''
and model ``B'', instead, have the same galaxy model but different ICM densities 
and velocities, resulting in an ICM kinetic pressure, $\rho_{ICM}U_G^2$, 10 times 
larger for model ``A1'' than for model ``B''. Finally, model ``Bq'' is identical to 
model ``B'', except that drop out is included, with $q=0.4$. 
The ICM density of models ``A'' corresponds to a fairly rich cluster,
while the density of model ``B'' is that of a poor cluster.
By comparing the 
results of the simulations between models, we attempt 
to derive the 
general properties of stripping dynamics and to see under which conditions
stripping is important.

\section{Hydrodynamic evolution}

As indicated in Table \ref{partable}, two different orbits and cluster 
models were considered. 
Fig. \ref{inflaw} shows, in the two cases, the orbits of the galaxy 
against the ICM background density, and the Mach number and the momentum 
flux of the corresponding flow upstream of the galaxy in the reference 
frame of the galaxy. 

\begin{figure*}[\protect{h!}]
\begin{center}
\begin{tabular}{cc}
\includegraphics[height=6.5cm,clip=]{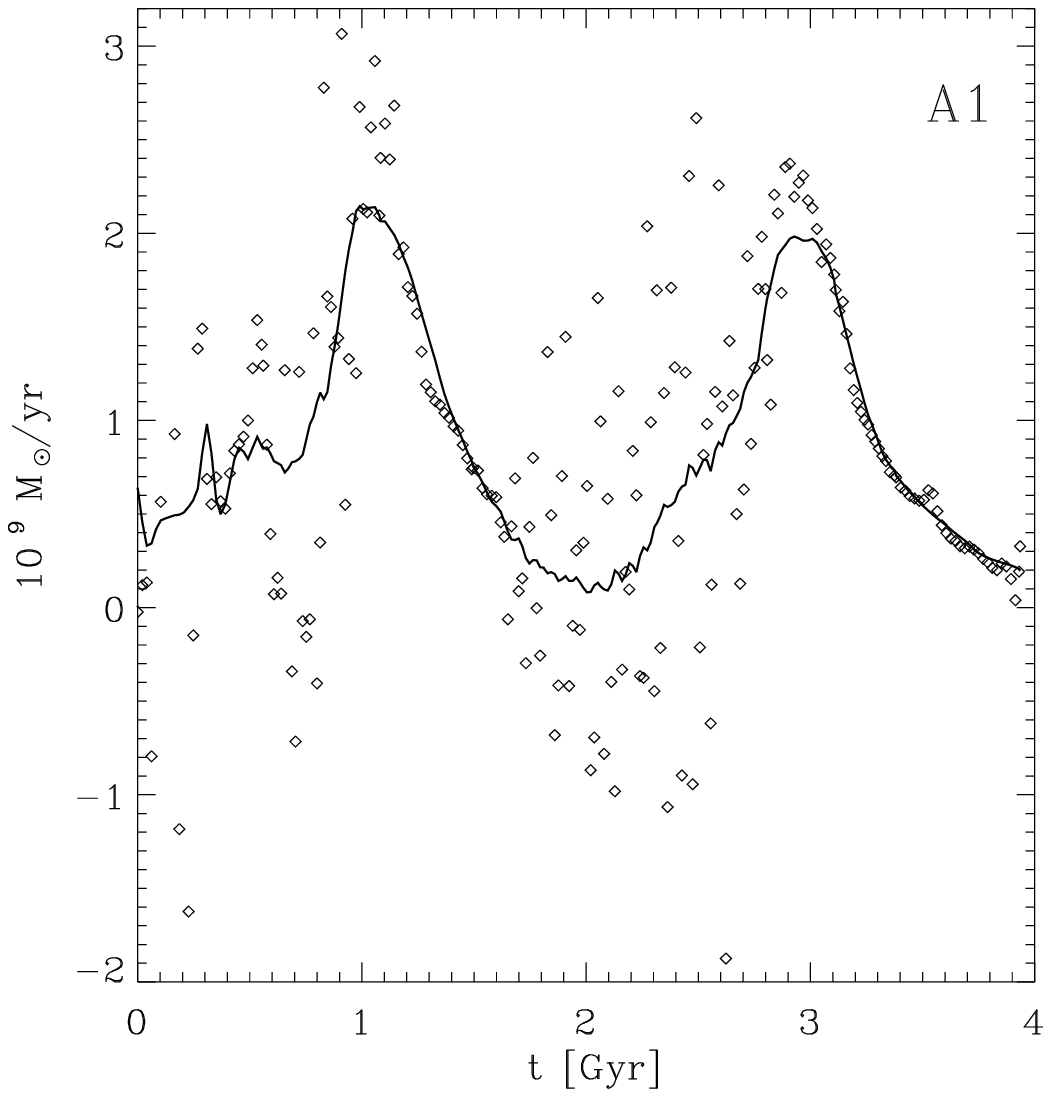}\put(-190.,90.){\rotatebox{90}{\scalebox{.6}{\msolar/yr}}}
&
\includegraphics[height=6.5cm,clip=]{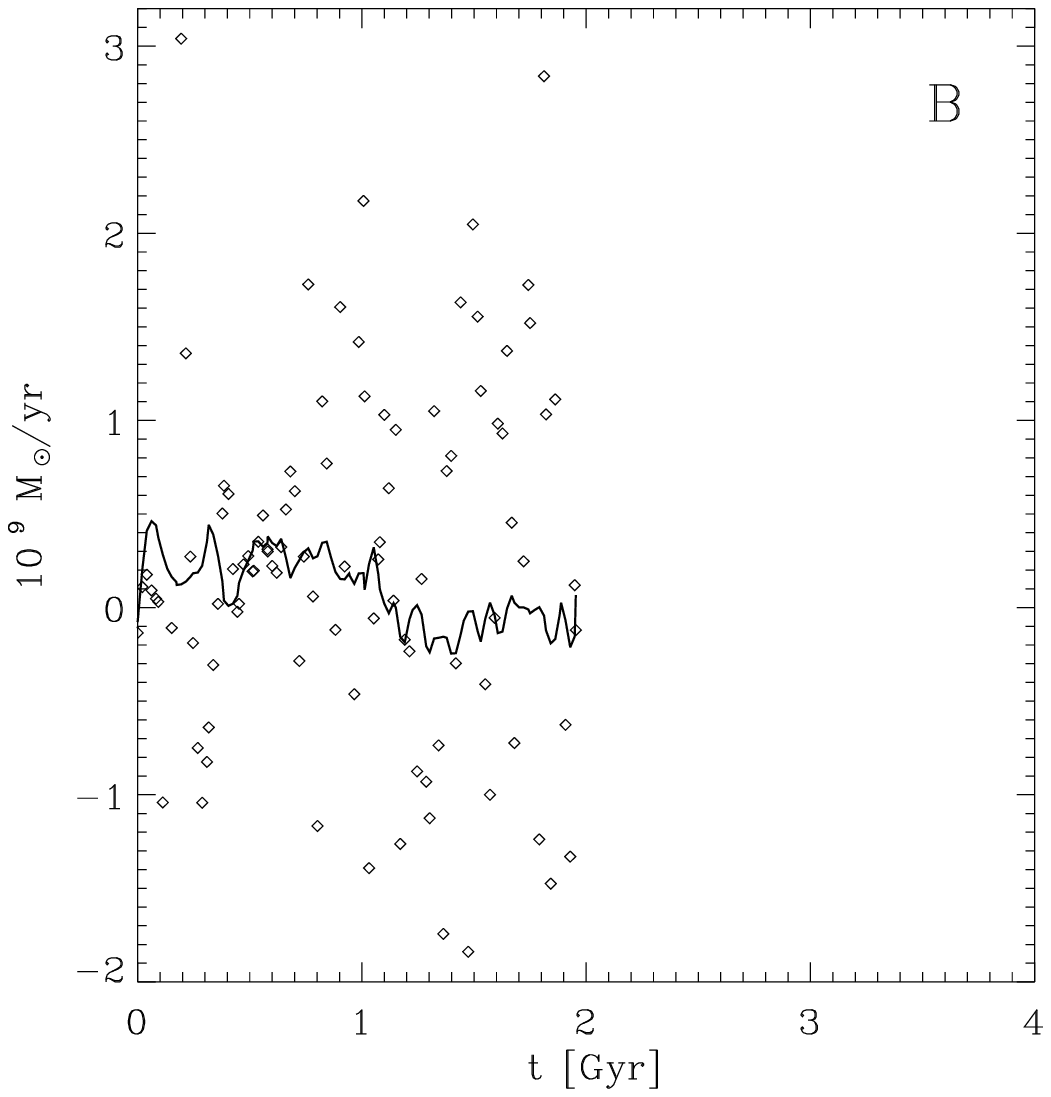}\put(-190.,90.){\rotatebox{90}{\scalebox{.6}{\msolar/yr}}}
\\
\includegraphics[height=6.5cm,clip=]{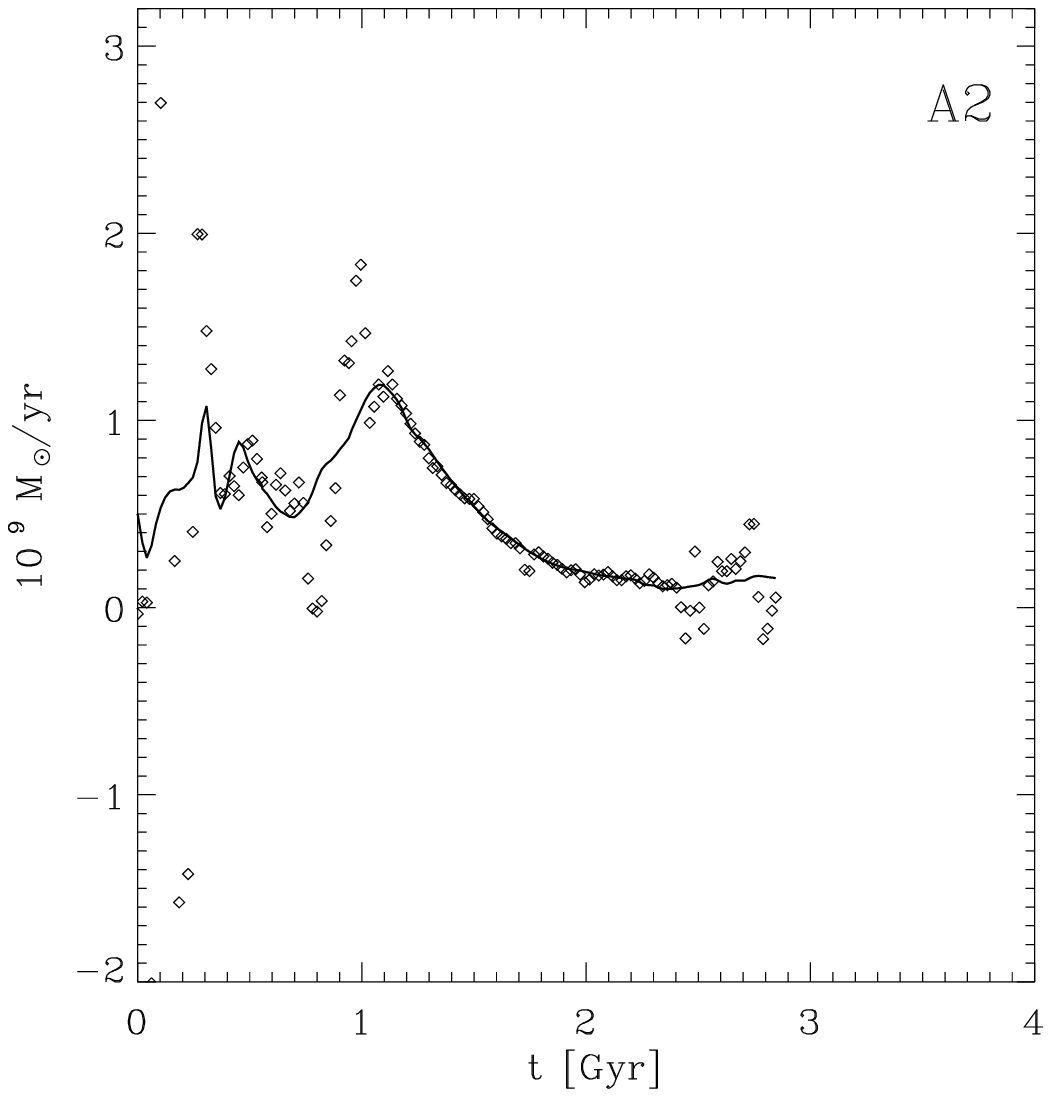}\put(-190.,90.){\rotatebox{90}{\scalebox{.6}{\msolar/yr}}}
&
\includegraphics[height=6.5cm,clip=]{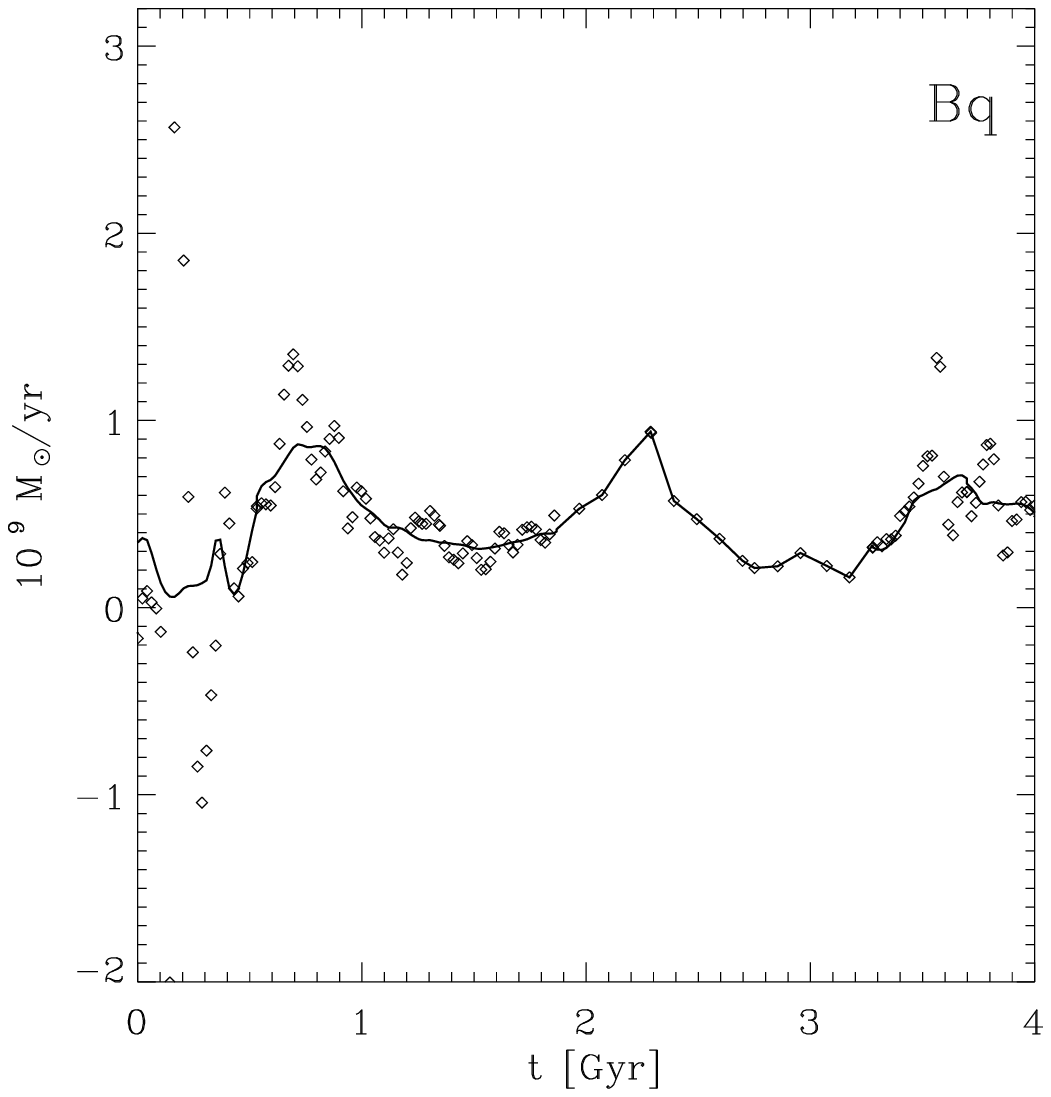}\put(-190.,90.){\rotatebox{90}{\scalebox{.6}{\msolar/yr}}}
\end{tabular}
\end{center}
\caption{\small Net ISM mass-loss rate as a function of time for the
four models presented, as labelled in the upper-right corner of each
panel. The dots represent the ratios of the increments in gas mass to 
the increments in time, for the gas contained within a radius $r_*$ of
the galaxy. The time resolution of the simulation data output rate was 
typically 15-20 Myr, or 100 time steps. The contribution of mass
injection has been added to the mass loss; for model B$q$, the gas
mass deposited as ``drop out'' has been subtracted from it. The solid
lines represent the mass-loss rate derived from smoothing over 10 to
20 crossing times. For model B$q$
%
the data 
between 1.9 and 3.3 Gyr 
have been 
obtained from the less frequently-sampled 3-D output files; no smoothing has 
been applied in that region to obtain the solid curve.\label{massloss}}
\end{figure*}

In Fig. \ref{massloss} 
the net mass-loss rates, corrected for mass injection 
and drop out, are given. In the absence of hydrodynamic interaction the 
galaxy would accumulate the gas shed by its stars, either in gaseous form 
or in the form of drop outs. This situation is referred to as zero
mass loss (or gas accumulation). 
When gas is lost from the galaxy, the mass-loss rate is defined as 
\be 
\dot M_{ISM}=\int{-\dot\rho+\dot\rho_{inj}-\dot\rho_{do}\,d^3x}
\ee
where $\dot\rho$ is the time derivative of the gas density, 
$\dot\rho_{inj}$ and $\dot\rho_{do}$ are the mass injection 
and the mass drop-out terms as defined in Eqs.(19) and (20), 
and the domain of integration extends over a suitably 
defined portion of space associated to the galaxy. Unless stated
otherwise, we will usually refer to the volume enclosed within 
a distance $r_*$ from the galaxy centre. Except for the first 
large stripping event, this estimate turns out to be very close to
that obtained when summing over all computational cells in which the
gas has a negative total energy density in the galaxy's reference 
frame, i.e. $\rho u^2/2 + \rho\Psi_G + 3nT/2<0$.

The panels corresponding to models A1 and B show that the 
hydrodynamic evolution entails not only gas loss, but also more 
or less prolonged phases of gas accumulation or even accretion (negative 
mass loss). 

A distinctive property of the mass-loss rates $\dot M_c(t)$ is their 
``noisiness'', especially when gas accumulation/accretion takes place. 
The dots in the plots of Fig. \ref{massloss} correspond to simulation 
data sampled at intervals comparable to one crossing time. In model A2, 
the galaxy is nearly bare of ISM, while in model B$q$ there is extra 
dissipation due to mass drop out, which in our prescription acts in a 
feedback fashion. In such cases the scatter in $\dot M_c(t)$ is 
reduced. 

The same reduction in scatter is observed in the stages of model A1 following 
pericentre passages. In those stages, mass loss is essentially determined by 
force balance. The kinetic pressure on the ISM gradually falls below the 
gravitational restoring force over a wider and wider region of the galaxy 
causing the observed decrease in the mass-loss rate. Force balance dominates 
most of the gas dynamics in model A2, where the galaxy has only about half the 
gravitational mass of the other three models, and the gas is not observed 
to cool significantly. However, for this model some scatter with short 
accretion stages can be seen at later times, a signature that a different 
dynamic regime sets in. 

The kinetic pressure force is the main gas dynamic agent when the upstream 
flow is supersonic and stripping is efficient. When a distribution of 
relatively dense ISM is present or is allowed to form, the effects of 
dynamic self shielding, of the spreading of vorticity from the ISM-ICM 
interface, and of gas cooling are important. In model B, gas cooling 
significantly increases the ``survival rate'' of the ISM distribution during 
the supersonic stages of the flow corresponding to orbital pericentre 
approach. The result is that a net average cooling-driven gas accretion 
occurs for this model within a radius of 16 kpc from the galaxy centre at 
later times.
The non-linear development of 
Kelvin-Helmholtz modes at the ISM-ICM interface near the galaxy centre, 
with time scales of less than a crossing time (in subsonic motion), is 
the cause of the chaotic alternation of gas loss and gas accretion in the 
outer parts of the orbit. This proceeds until the growing kinetic pressure 
significantly disturbs the equilibrium of forces again. We discuss the two main 
dynamic regimes, supersonic ram-pressure stripping and subsonic K-H 
dynamics in the following two section of this paper. 

\subsection{Supersonic stripping}

\begin{figure*}[\protect{h!}]
\begin{center}
\includegraphics[width=6.2cm]{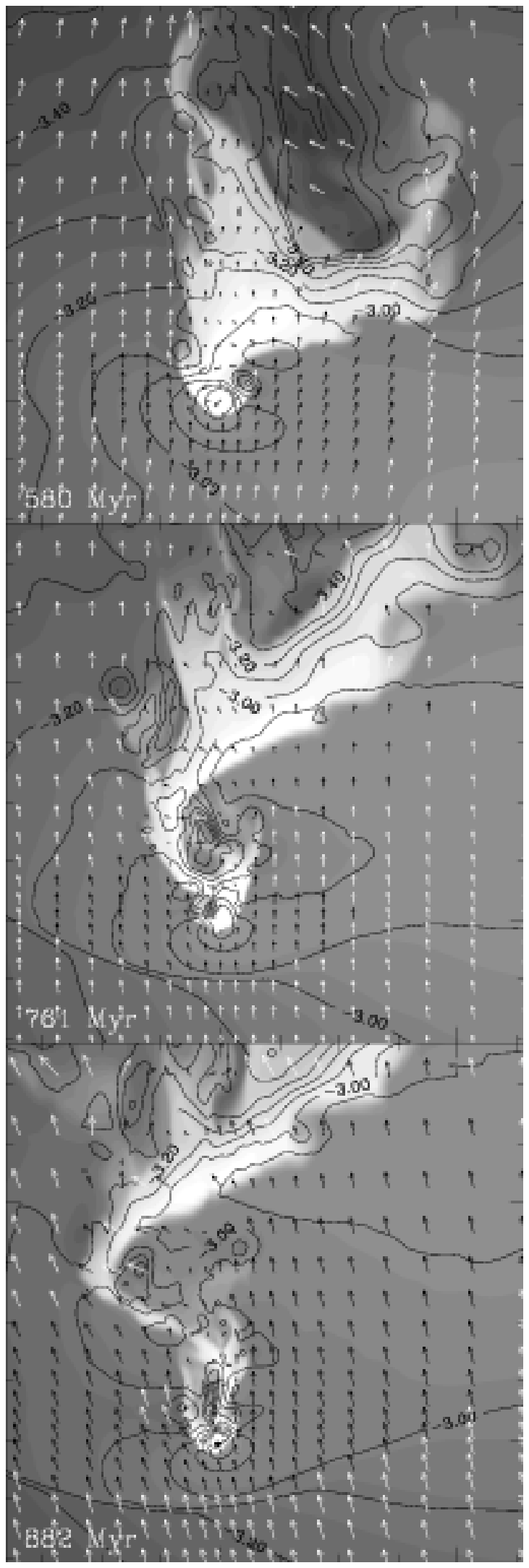}
\includegraphics[width=6.2cm]{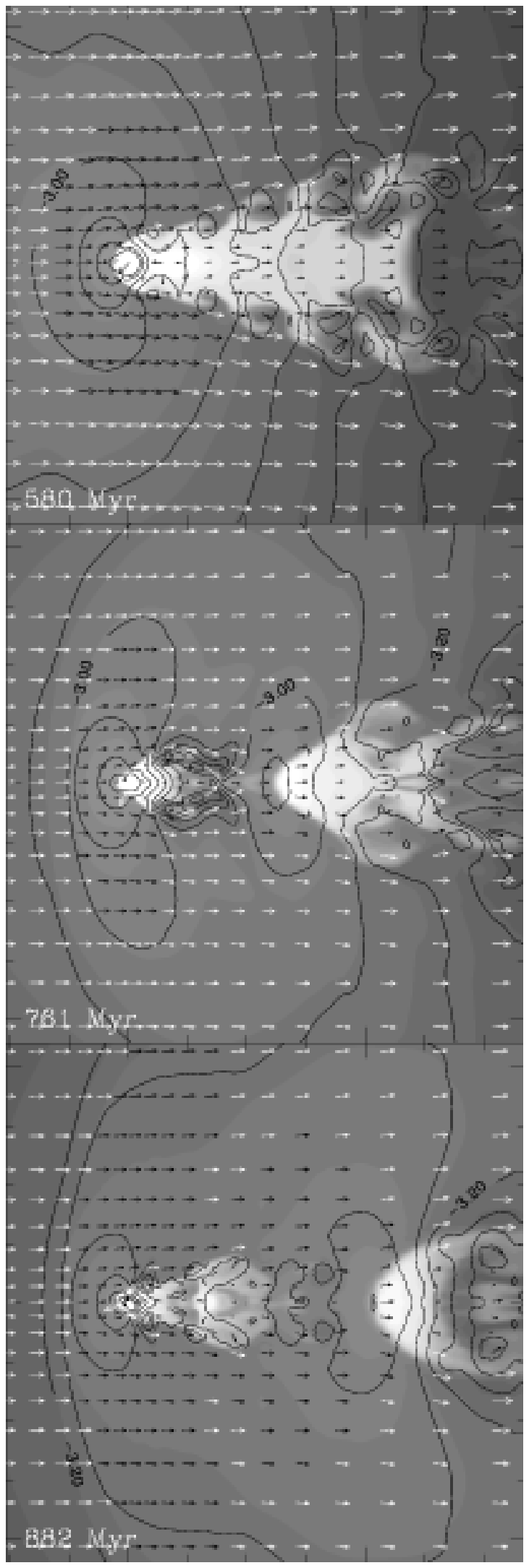}
\vspace{-.25cm}

\includegraphics[width=5.8cm]{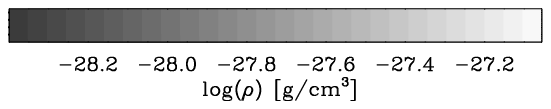}
\end{center}
\vspace{-.25cm}

\caption{\small Gas density (grey scale at the bottom), pressure (black 
contour lines, logarithmically spaced from log$(p)$=-3.6 to log$(p)$=-2.6 
in units of $10^{-24}$keV/cm$^3$), and Mach vectors 
(arrows, drawn in white when Ma$>1$, and black otherwise) 
of model B. Coordinates are given in kpc. 
Numbers on the bottom left of each panel give elapsed time in 
Myr. A section in the plane of the orbit (panels on the left) and one in an 
orthogonal plane (panels on the right) are given for each time. 
\label{machex}}
\end{figure*}

Gas-mass loss is largest in the part of the galaxy's orbit near the pericentre 
where the kinetic force is largest. Its dynamics is regulated by the pressure 
distribution around the ISM produced by the supersonic flow. 
Rarefaction 
fans are formed at the rear of the galaxy at the same time as pressure builds 
up in front of it behind the bow shock. 
The presence of the bow shock itself contributes to the lateral pressure 
confinement of the ISM, while the ISM accelerates and expands where rarefaction 
fans form in the streaming ICM. As a result, the ISM is initially pushed back 
bodily, stretched and fanned out in the direction of motion. Figure \ref{machex} 
shows the evolution of model B during this phase. In contrast to the case of 
a body with a hard surface, no primary trailing shock is formed, because the ISM 
is free to expand. In time, K-H wiggles develop on the ISM-ICM interface, which 
produce secondary shocks and secondary rarefaction fans on the sides downstream 
of the galaxy centre. Irregular extensions or ``tongues'' of ISM form, which grow 
or stretch until they break away. The time scale of the 
whole process is intermediate between the dynamical crossing time and 
the sound crossing time 
for the cooler gas ($c_s\simeq 400$ km/s, as 
compared to $U_G\simeq1200$ km/s). 

The amount of ISM actually lost from the galaxy's potential well depends on 
the kinetic pressure or momentum flux $\rho_{ICM}U_G^2$ and on how long it 
lasts. Stripping is boosted by rapidly growing K-H modes, but it is reduced 
by gas cooling and by counter streaming towards the centre of the galaxy from the 
rear (see also Nittmann et al. 1982). 
Condition (\ref{spannometrica}) strictly applies only if $\mach\gg1$, since 
in that case one can neglect pressure forces acting on the ISM. For mildly 
transonic stripping, Eq. (\ref{spannometrica}) is a sufficient condition 
for complete stripping only if it is valid in the centre of the galaxy and if 
it is satisfied over an ISM crossing time. 
If it is valid only outside a particular cylindrical radius, 
although that region will be depleted, it does not imply that all the ISM 
associated to it, which joins in the trail, will be lost in a dynamic time. 

%
Initially, the momentum acquired by the ISM is spent in climbing up 
the galaxy's potential well, so that for some time the ISM 
approximately co-moves with the galaxy. As soon as it is displaced and 
decelerated, it starts falling towards the cluster centre, since given a 
density ratio $T_{ICM}/T_{ISM}\sim5$ its buoyancy in the ICM is small. 
The cluster gravitational force (or tidal force in the galaxy's reference 
frame) is not small compared to the kinetic pressure force. 
A second shock is formed in front of the part of stripped, flowing material 
which is on the same side as the cluster centre 
with respect to the galaxy. The pressure built up behind the second shock 
in turn displaces the ISM upstream. This produces the ``S''-shape of the 
trail visible in Fig. \ref{machex}. The ISM nearer to the galaxy is subject to 
a pressure due to the large net momentum with respect to the galactic 
centre, which effectively ``breaks'' the trail and allows a flow of ISM 
towards the galactic centre. At the same time, other gas is stripped from 
the galaxy near the centre due to the growth of K-H wiggles, and the process 
may repeat on a smaller spatial scale. 

In model B, at this point the galaxy is slowing down moving outwards from 
the cluster centre. No further stripping occurs after a second, minor 
event, and in the centre most of the ISM is retained. Moreover, the 
compression before the formation of the bow shock and after the 
formation of the lateral shock causes a reduction of the central cooling 
time and a net mass accretion in the centre. Despite ram pressure, a central 
cooling accretion flow is established. A similar situation obtains for model 
B$q$, where, however, significant mass loss continues also during subsonic 
motion due to the reduced density contrast. In models A1 and A2 kinetic 
pressure is large and prolonged enough to stop accretion and cause again 
heavy stripping. The rate of accumulation of ISM in the subsonic part 
of the orbit is crucial to determine the gas content at later stages.

\subsection{Subsonic stripping flow}

The growth of the ISM halo while the galaxy is moving towards its 
apocentre is regulated by gas accumulation via stellar gas injection. It 
increases the ISM density gradient near the centre and thus inhibits the 
linear growth of K-H modes. Balance is thus shifted toward gas accumulation 
at the ISM-ICM boundary, which proceeds faster than the entrainment of 
streaming gas. The boundary of the ISM halo, which is marked by a sharp 
density gradient, moves outwards until a point is reached where K-H stripping 
balances, on average, mass accumulation. 
Vortices are mainly created by non-linear K-H modes and, by growing in 
size, they entrain additional gas, until they are either `washed' downstream 
or dissipated near the galaxy centre. The formation, advection and 
eventually dissipation of vortex structures thus traces events of abrupt 
loss or of more gradual capture of gas in the galactic gravitational 
potential well, which is related to significant mixing between the ISM 
and the ICM within the galaxy. 

\begin{figure}[\protect{th!}]
\begin{center}
\includegraphics[width=14.cm]{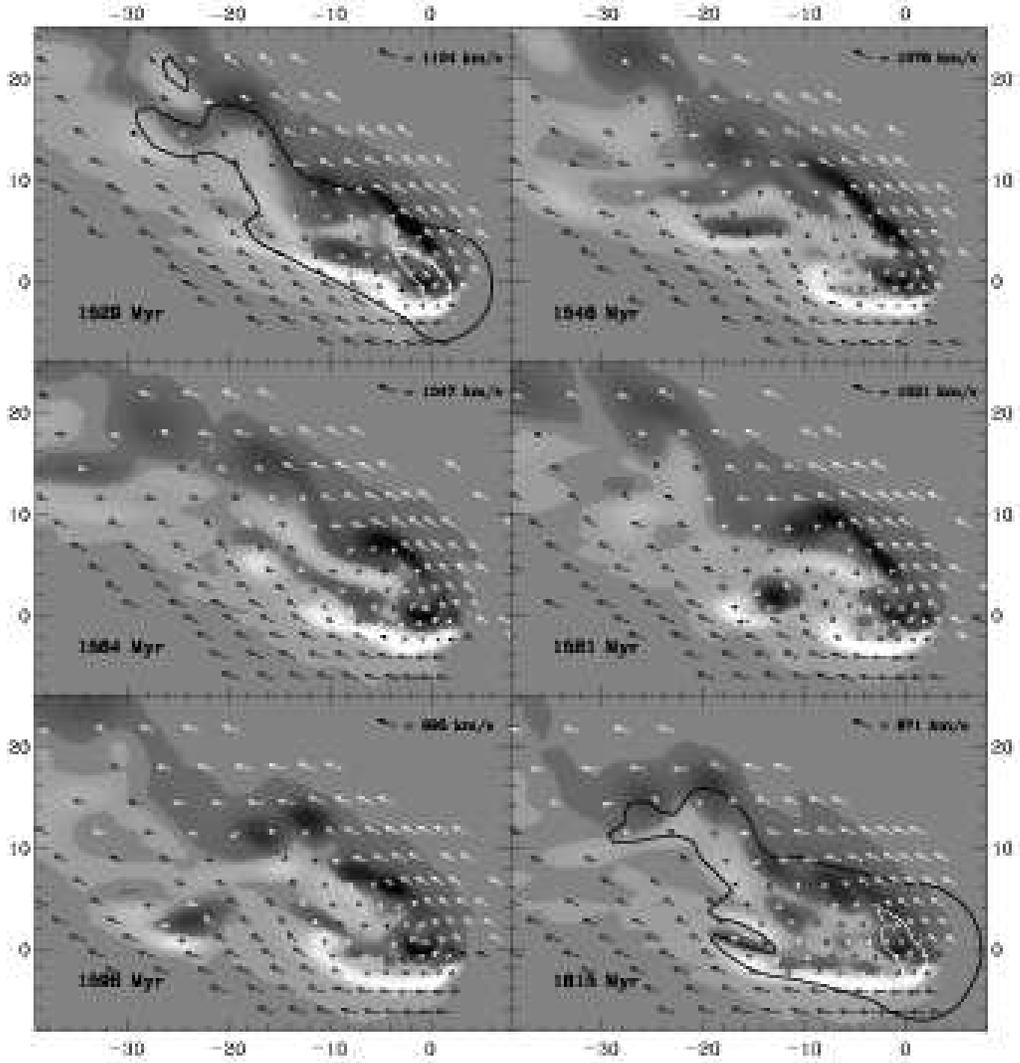}
\end{center}
\caption{\small Velocity and vorticity field in the equatorial plane at 6 
different times, at intervals of 17 Myr, for model A2. Arrows represent 
velocity vectors in the plane, 
while the background grey scale represents the corresponding z component of the 
vorticity field, from -0.61 (white) to 0.61 (black) Myr$^{-1}$. 
The time is given 
in each panel on the bottom left, the free-stream velocity of the ICM in the 
upper right corner. Lengths are given in kpc. 
The black contour lines in the first and in 
the last panel enclose the gas with negative binding energy density (kinetic 
plus potential); the white contours enclose the gas having negative total 
(binding plus thermal) energy density. \label{vort2} }
\end{figure}

\begin{figure}[\protect{h!}]
\begin{center}
\begin{tabular}{cc}
\includegraphics[width=7.cm,clip]{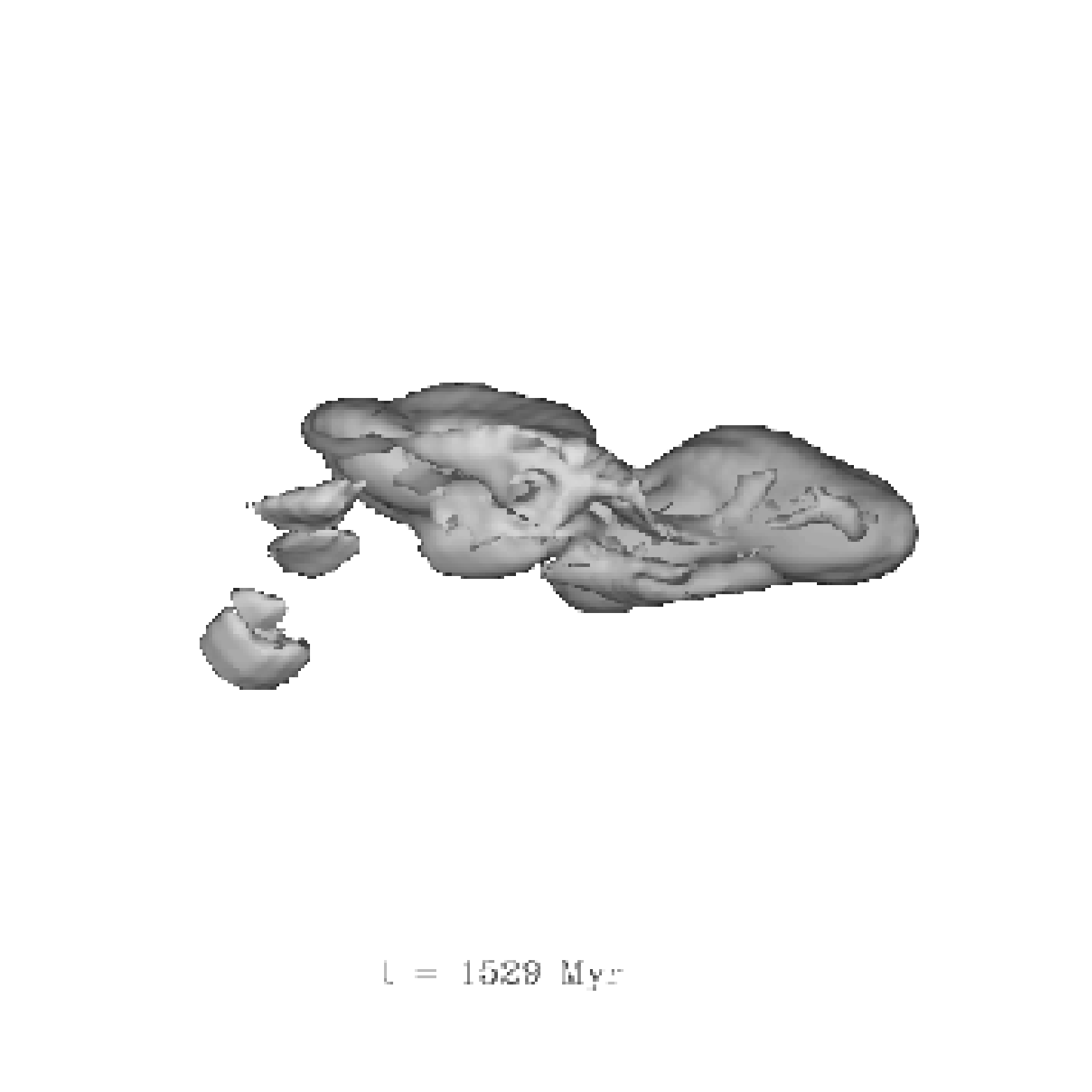}
\includegraphics[width=7.cm,clip]{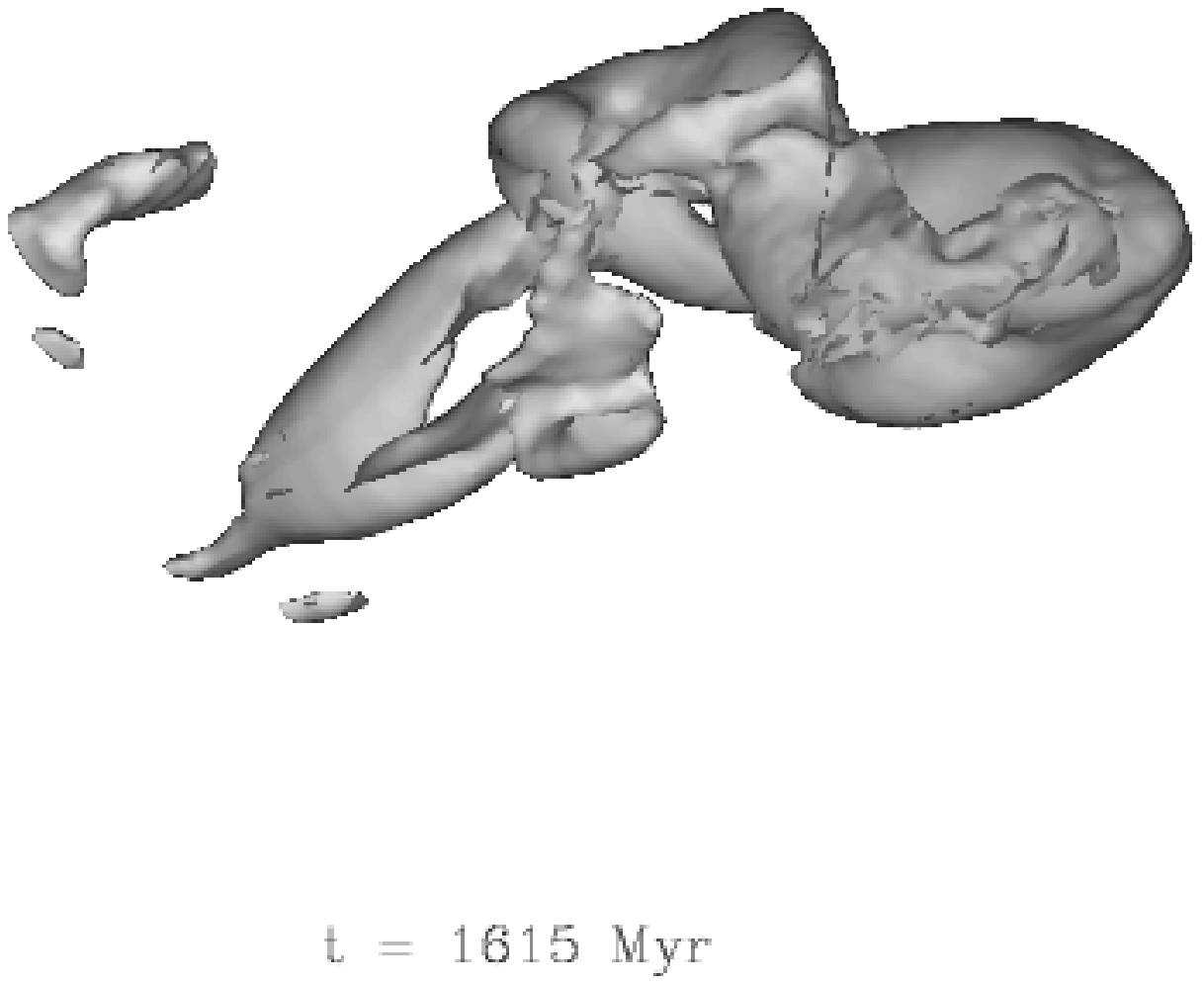}
\end{tabular}
\end{center}
\caption{\small Three-dimensional view of the vorticity field for model A2. 
A surface of constant vorticity ($|\omega|=0.0165$ Myr$^{-1}$) is shown for 
two different times (as labelled). The size of the ring structure in the 
upper part of the right-hand side figure is about 5 kpc.\label{vort3}}
\end{figure}

The velocity and vorticity fields during subsonic motion for model A2 are 
represented in Figs. \ref{vort2} and \ref{vort3}. In Fig. 
\ref{vort2}, a cut in the equatorial plane of model A2 is shown, where 
the evolution and advection of vortices can be distinguished, while ISM is 
accumulating within the galaxy. In the 
present context, generation of vorticity is exclusively connected to 
stripping. Vorticity spreads as the size of 
the region of bound gas (black contour in Fig.~4) grows. 
Near the centre of the galaxy, the gas keeps concentrating, and smaller 
vortices develop. The complicated three-dimensional structure of the vorticity 
field, which tends to be organized in trailing filaments and rings as the 
gas-mass loss from the galaxy decreases, is visible in Fig. \ref{vort3}. 
While stripping is still substantial (t=1529 Myr, left, corresponding to 
the top left panel of Fig. \ref{vort2}), the vorticity field is more
streamlined  
and distributed according to the stripped gas. Later, at t=1615 Myr, less 
regular structures appear, with an alternate shedding of vortex rings analogous 
to that occurring 
in 3-D aerodynamic flows past blunt bodies (Goldstein 1938, \S 250, 
p.577-579; Perry \& Lim 1978). 

Although
the flow is only mildly subsonic, its character is very different from that 
of a case of ram-pressure stripping. The motion of the galaxy is communicated 
to an extended region of the ICM and kinetic pressure on the ISM is small. 
It is at this stage the necessity of a very wide computational grid (a 
half cube of 700 kpc size) becomes apparent. 
Comparison 
between the usual and the high-resolution (on a much smaller grid) 
computations we performed for cases A2 and B indicated consistency for the 
stripping rates. 


Near the centre of the galaxy the ISM is rotating with angular momentum 
roughly aligned to the galaxy's orbital angular momentum. The gas accreted 
has a net angular momentum in that direction. The flow round the galaxy in 
the equatorial plane 
has
a non-zero 
circulation corresponding to an enhanced mass loss on the side of the galaxy 
opposite to the cluster centre (lower panel in Fig.  
\ref{vort2}), and enhanced accretion on the other side. In an orthogonal plane 
a similar situation may be attained but in certain regions and on short 
intervals of time only, the sign of the circulation changing frequently. 
The accretion region is much larger in model A1 than in model A2, where the 
flow pattern is even more complex and non-planar.


\begin{figure*}[hb!]
\begin{center}
\includegraphics[height=10.cm,width=13.cm]{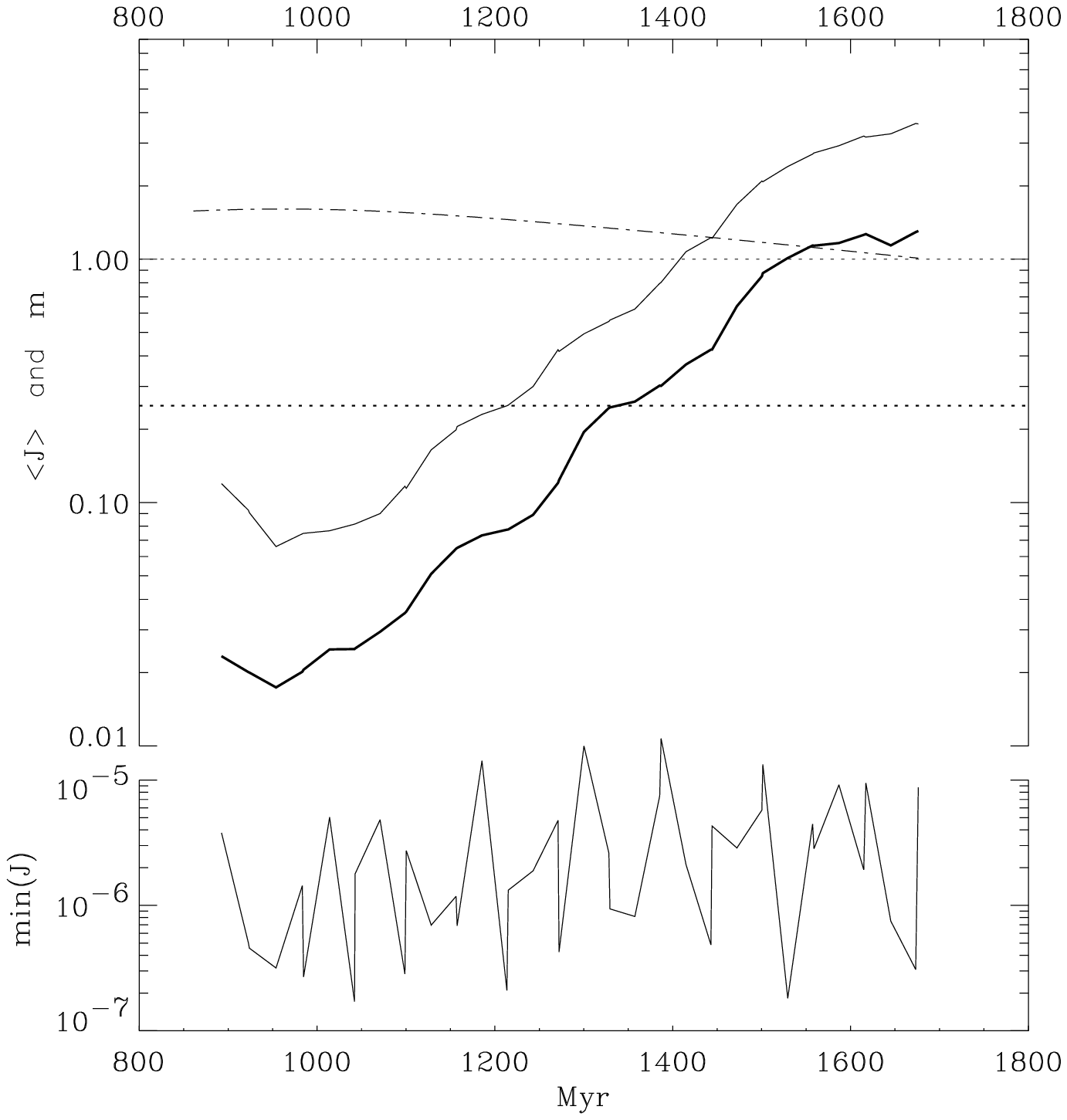}
\end{center}
\caption{\small Richardson number $J$ (solid lines) and Mach number
(dash-dotted line) 
as a function of time for model A2 (see Fig. \ref{vort2}). The thick
solid line shows the value of the average $J$ over the high-vorticity region 
evidenced in Fig. \ref{vort2} (defined to have a vorticity 
$|\omega|\ge188$ km s$^{-1}$ kpc$^{-1}$). 
The thin solid line is the median of $J$ over that same region. The bottom
part of the plot shows the minimum value of $J$ in the region considered. 
\label{richfig}}
\end{figure*}

An indication for the effectiveness of the K-H instability in determining 
the flow configuration is given by the value of the Richardson number 
(Figure \ref{richfig}), defined by 
\be 
J \equiv \frac{\nabla\psi\cdot({\boldmath \omega}\times\bu)\:
               \nabla\ln(\rho)\cdot({\boldmath \omega}\times\bu) }
              {\omega^2 |{\boldmath\omega}\times\bu|^2} \, . \label{defJ}
\ee
Incompressible K-H modes are stabilized when $J\gsim1/4$ locally 
(Chandrasekhar 1961, \S 103, p 491). Conversely, growing K-H modes are 
likely to exist as long as $J<1/4$, with wavelengths of the order of the 
scale given by the logarithmic density gradient (neglecting tidal forces: 
op. cit., \S 104). Compressibility, i.e. a free-stream Mach number Ma $\sim1$, 
generally has a stabilizing effect. In our models we observe that the `noisy' 
accretion phases correspond to average values of $J>1/4$, indicating that K-H 
modes are saturated. Locally, however, low values of $J$ are attained at all 
times, mainly as a result of mass injection via stellar mass loss.

\subsection{Stripping and drop out\label{strip-and-drop}}

\begin{figure*}[t!]
\begin{center}
\includegraphics[width=14.cm]{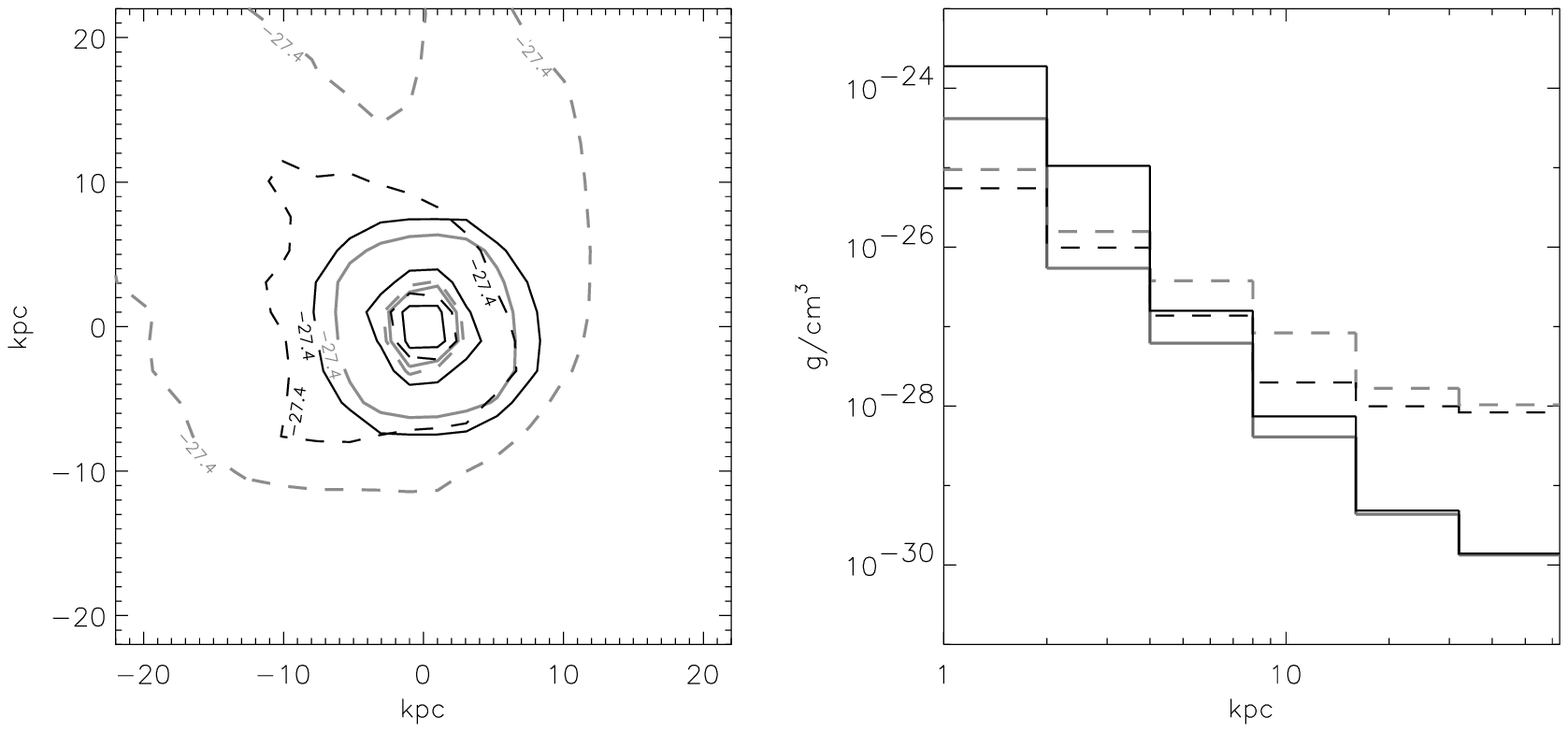}
\end{center}
\caption{\small  {\it Left:} Contours in the equatorial plane for the densities 
$\rho_{do}$ and $\rho$ (solid and dashed lines, respectively) for run B$q$, 
where $\rho_{do}$ represents the accumulated mass deposition per cm$^3$. 
The gas is assumed to ``freeze out'' at the location of the reference 
system attached to the galaxy where it is lost from the normal gas phase.
The distributions are given at $t=976$ Myr (grey 
lines) and $t=4055$ Myr (black lines), respectively. The contour levels 
of the decimal logarithm of the density in g/cm$^3$ are -27.4, -25.7 and 
-24.
{\it Right:} radial profiles from the spatial density distributions of 
the normal ISM and of that in drop-out form. Different line styles have 
the same meaning as the contour lines in the left panel. 
\label{drop-out}}
\end{figure*}

Model B$q$ displays important deviations 
from the dynamics outlined above. From comparison of models B and B$q$ 
in Fig. \ref{massloss}, 
it is seen that drop out 
significantly affects the ISM content of the galaxy, and its dynamics. 
In general, a more nearly steady, self-regulated flow is achieved, having a 
lower density contrast between the gas near the centre of the galaxy and 
the ICM. Stripping is more continuous, but larger ISM mass-loss rates are 
obtained. The gas temperature undergoes smaller variations than in case 
B, and is generally higher as cold gas is dumped via drop out. 

The drop-out rate is especially large when the ISM is bodily compressed 
while the galaxy is accelerated towards the pericentre. The reduced density 
contrast favours mass loss at intermediate radii in the subsonic stage. This 
further reduces the density contrast. As a consequence, less ISM is retained by
the galaxy, either in gaseous or in drop-out form. 
Our quantitative results are sensitive to prescription (\ref{dropout}) and 
to the value of $q$ adopted. However, 
the qualitative conclusion that drop out favours mass loss is 
robust.


In the specific case of model ``B$q$'', with $q=0.4$, drop out 
occurs mostly in the centre of the galaxy, 
at such rate that after 4 Gyr the accumulated drop-out 
mass is $\sim$14 times larger than the ISM gas mass within a galactic 
half-luminosity radius ($r_*=16$ kpc). From Fig. \ref{drop-out} one can see
that the mass distribution of gas dropped out of the flow keeps steepening as 
time passes. For $t>1$ Gyr, drop out is essentially confined to the central 
20-30 kpc,  with more of half the drop-out mass contained within 8 kpc. This 
regulates the radial profile of the ISM density, which, in contrast to what is 
observed in the other models, does not steepen as time passes. On the 
contrary, the central density even decreases. 

In the subsonic phase, while a smaller density gradient allows a faster 
growth of K-H modes, the mass-loss term in Eq. (\ref{mass}) acts as 
an effective viscosity in the flow (Portnoy et al. 1993, Kritsuk et al.
1998). The 
K-H modes grow faster but saturate 
earlier.
The saturated modes are associated with smaller mass flows 
also due to the lower ISM density. Thus, the mass-loss process is much 
smoother than in the non-drop-out models. 

\subsection{Secular evolution\label{secular}}

The gas content of the model galaxies oscillates over a time equal to 
the radial period of the galaxy's orbit, and heavy stripping alternates 
with replenishment and even cooling-driven accretion. The evolution, 
however, is not cyclic over the computed time interval, but shows rather 
a long-term tendency towards either complete cyclic stripping or a 
permanent cooling accretion flow onto the galaxy centre. In other words, 
no equilibrium between cooling and stripping and/or heating is achieved. 
Rather, in the central parts of the galaxy, one of the two processes 
eventually becomes ineffective. This is related mainly to the character 
of optically thin thermal radiative cooling (see also Kritsuk 1992), and 
to a lesser extent to the diminishing rate of gas replenishment from 
stars (Eq. (\ref{Renzini})). 

In model B, cooling is sufficient to start a central cooling accretion 
flow (essentially from the rear side) during the supersonic phase, gas 
accumulation/accretion is effective afterwards, and a further pericentre passage 
does not affect the ISM anymore. 
In model A2, ram pressure is large enough to strip the galaxy completely (if we 
consider only the gas having a negative total energy density). Model A1 
is intermediate between the two, in that stripping affects the ISM also in 
the centre, but does not stop gas cooling which results in the formation of 
a central accretion flow in the subsonic part of the orbit. However, the
next pericentre passage, 3 Gyr after ``infall'', causes complete stripping. 

\begin{figure*}[ht!]
\begin{center}
\includegraphics[width=14cm] {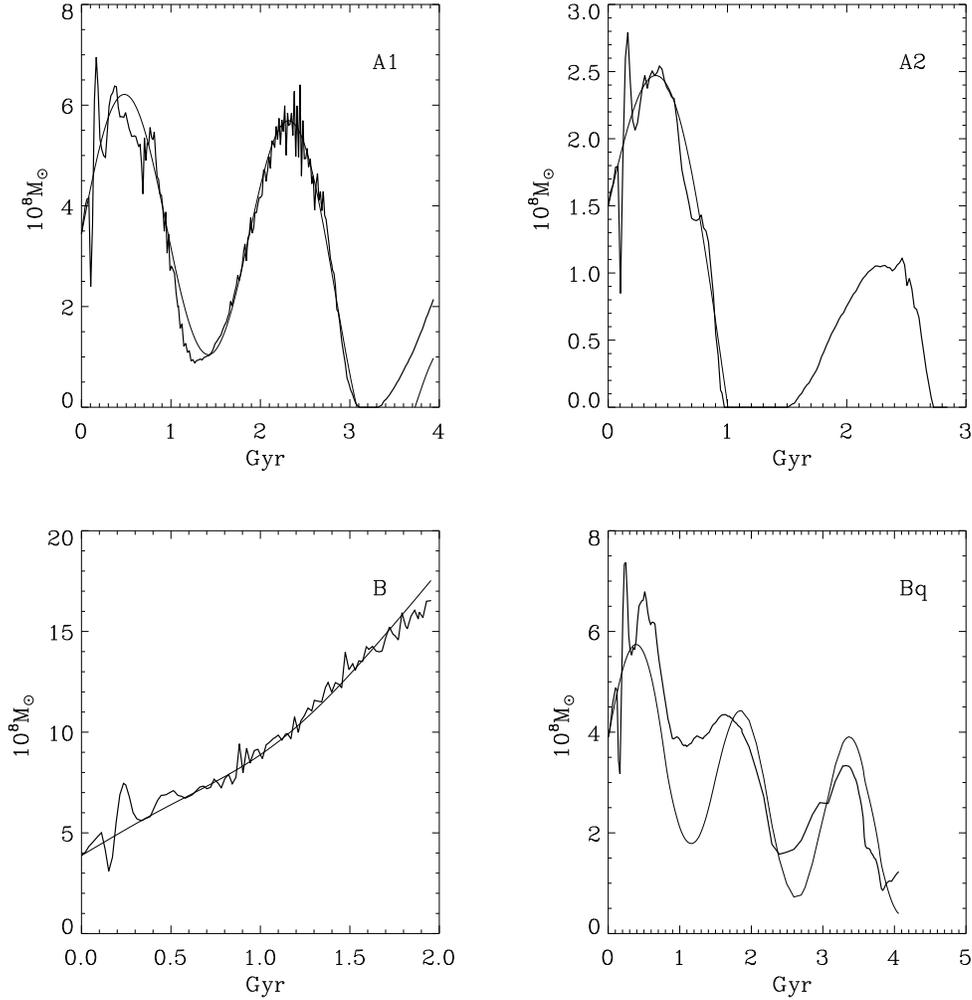}
\end{center}
\caption {\hspace{.5cm} Secular evolution of the ISM content of the model 
elliptical galaxies. The thick lines represent for each model the total 
mass of the gas with negative total energy density in the galaxy's reference 
frame. 
The thin smoother curves 
are fitted as described in the text (Eq. (\ref{stimasec})).
\label{secularfig}}
\end{figure*}

We may attempt to quantify the long-term behaviour with an evolution 
equation for the ISM mass of the form 
\be 
  \frac{dM_{ISM}}{dt} \simeq 
          \frac{\chi M_{ISM}}{\tau_{c0}} - 
          \frac{\eta M_{ISM}\!(0)^{1+b}}{\tau_{orb}M_{ISM}^b}+
          \alpha_*M_*\left(\frac{\tau_*}{t}\right)^{1.3}
\, , \ee
where $\tau_*$ is the age of the stellar population of the galaxy, $\tau_{c0}$ 
the initial central cooling time, $\tau_{orb}$ the radial orbital period, 
$\alpha_*M_*$ the initial mass-injection rate, and $M_{ISM}(0)$ the initial 
ISM mass (intended here, as before, the total mass of gas with negative 
total energy density). The first term on the r.h.s. is related to cooling 
(with $\tau_c^{-1}\propto\rho$), the second to stripping, and the last 
is the mass injection. The mass dependence (or density dependence) 
through the parameter $b$ in the second term allows for the dependence 
of stripping efficiencies on the density contrast. If for simplicity we take 
$b=0$ and $\chi=1$, using $\dot M_{inj}\simeq M_{inj}(0)(1-1.3t/\tau_*)$ we 
obtain 
\bea
 \lefteqn{ M_{ISM}  
           \simeq M_{ISM}\!(0) + 1.3\frac{\tau_{c0}}{\tau_*}M_{inj}\!(t) + }  
\nonumber \\
 &&  \left[ M_{inj}\!(\tau_{c0})\left(1-0.65\frac{\tau_c}{\tau_*}\right) + 
     M_{ISM}\!(0)\left(1 - \eta \frac{\tau_{c0}}{\tau_{orb}} \right)\right]
     \left(\:{\rm e}^{t/\tau_{c0}}-1\right) 
\, . \label{stimasec}\eea
Here, $\eta$ is a ``fit'' parameter characterizing the efficiency of 
stripping. If we add an oscillatory component to this estimate with a 
period equal to $\tau_{orb}$, a phase shift of a quarter period with 
respect to the orbital velocity, and a best fit amplitude, we find the 
fitting curves plotted in Fig. \ref{secularfig} together with the simulation 
data. The fitting curves clearly make sense only as long as the 
mass is positive. The values of $\eta$ appearing in Eq. \ref{stimasec} 
and of the amplitude $A$ of the sinusoidal components are given in Table 
\ref{sectab}. 

\begin{table}
\renewcommand{\arraystretch}{1.2}
\begin{center}
\begin{tabular*}{14.cm}{@{\extracolsep{\fill}} l @{\hspace{2cm}} c c c c c}
 & \bf A1 & \bf A2 & \bf B  & \bf Bq \\ \hline
$M_T$         & 11  & 5.7 & 10  & 10  \\
  $\bar m$     & 1.23 & 1.23 & 0.95 & 0.95 \\
$\bar\sigma$ & 1.31 & 1.31 & 0.15 & 0.15 \\
      $q$            & 0    & 0    & 0    & 0.4  \\ \hline
\boldmath $\eta$     & 4.20 & 4.25 & 2.56 & 3.67 \\ 
\boldmath $A$        & 2.6  & 1.4  & 0.25 & 1.6  \\ \hline
\end{tabular*}
\end{center}
\caption{\small \label{sectab} Fit parameters for the long-term
evolution: total gravitational mass $M_T$ 
of the galaxy in units of $10^{11}\msolar$, time-averaged Mach number $\bar m$
of the galaxy velocity relative to the ICM,
time-averaged incoming momentum flux $\bar\sigma$ 
of the ICM in the frame 
of the moving galaxy in units of $10^{-11}$ g cm$^{-2}$ s$^{-1}$,
mass drop out parameter $q$, parameter characterizing the
efficiency of stripping $\eta$
and amplitude of sinusoidal components $A$.
}
\end{table}

The fit, although corrected for drop out, is poor for model B$q$. The value 
for $\eta$, though, reflects the highly increased efficiency of stripping 
for this model as compared to its non-drop-out counterpart, model B. 

\begin{figure*}[ht!]
\begin{center}
\includegraphics[width=14.cm, bb=56 250 480 550, clip]{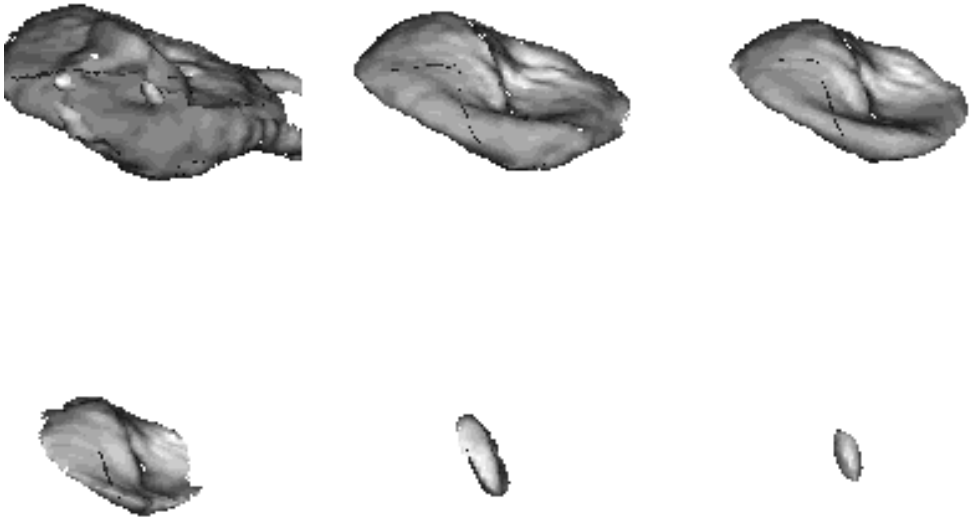}
\end{center}
\caption{\small Disk-like ISM density distribution in model B at $t=$1870 Myr. 
Shown are surfaces of constant density. The six density levels, increasing 
from left to right and from top to bottom, are 
8$\times10^{-28}$, 3.2$\times10^{-27}$, 1.28$\times10^{-26}$, 5.12$\times10^{-26}$, 
2.05$\times10^{-25}$, and 8.19$\times10^{-25}$ 
g/cm$^3$. On the bottom left the linear scale in kpc is given. 
The ICM streams by from the left. 
\label{disk}}
\end{figure*}

In model B, as a result of the angular momentum acquired during stripping, 
the ISM cools onto a large central disk in the plane of the galactic orbit 
with a diameter of about 20 kpc, as shown in Fig. \ref{disk}. The motion 
(essentially rotation) of the gas in the disk is supersonic, so that the disk 
is dynamically `cold' with a temperature of 0.01 keV. The disk is warped and 
presents spiral arms which join a central bar-like structure lying in the 
plane of the disk and oriented perpendicularly to the motion of the galaxy.
The gas flows towards the centre of the galaxy along the `bar', which 
is very cold (0.001 keV, the lower limit allowed for in the 
computation) at a radius of $\sim 5$ kpc, but it is warmer in the centre 
because of the compression caused by infall. While some or all of the central 
infall may be artificial due to the resolution limits of the computation, 
the formation of the large disk structure seems inescapable. 

\section{Mass Loss and X-ray Emission}

\begin{figure*}[\protect{th!}]
\begin{center}
\includegraphics[width=14.cm, bb=107 361 473 738, clip]{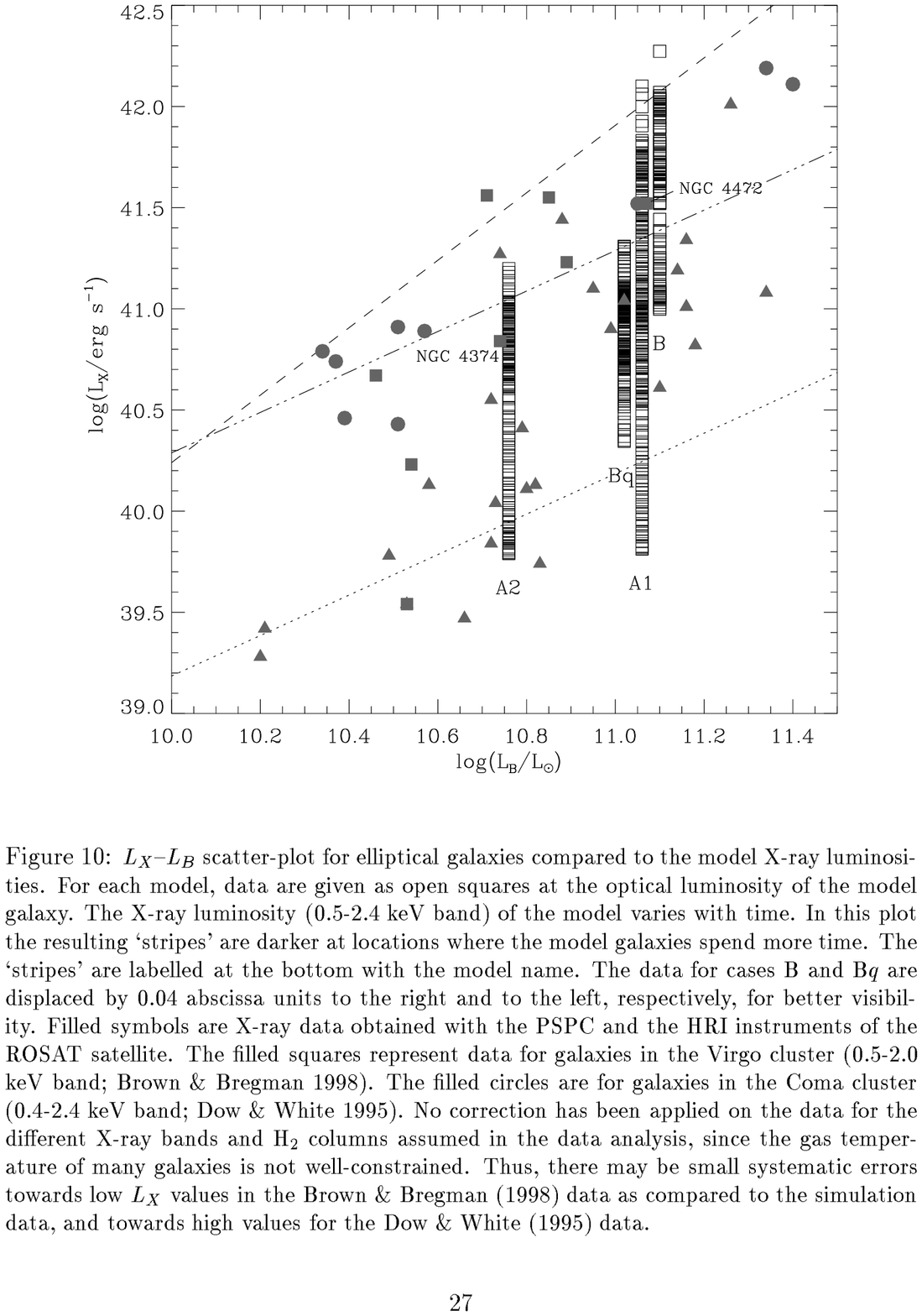}
\end{center}
\caption{\small $L_X$--$L_B$ scatter plot for elliptical galaxies compared to 
the model X-ray luminosities. For each model, data are 
given as open squares at the optical luminosity of the model galaxy. The 
X-ray luminosity (0.5-2.4 keV band) of the model varies with time. In this 
plot the resulting `stripes' are darker at locations where the model 
galaxies spend more time. The `stripes' are 
labelled at the bottom with the model name. The data for cases B and B$q$ are 
displaced 
by 0.04 abscissa units to the right and to the left, respectively, for 
better visibility. Filled symbols are X-ray data obtained with the PSPC 
and the HRI instruments of the ROSAT satellite. The filled squares represent 
data for galaxies in the Virgo cluster 
(0.5-2.0 keV band; Brown \& Bregman 1998). The 
filled circles are for galaxies in the Coma cluster (0.4-2.4 keV band; 
Dow \& White 1995).
No correction has been applied on the data for the different X-ray bands and 
H$_2$ columns assumed in the data analysis, 
since the gas temperature of many galaxies is not 
well constrained. Thus, there may be small systematic errors towards low 
$L_X$ values in the Brown \& Bregman (1998) data as compared to the simulation 
data, and towards high values for the Dow \& White (1995) data. 
\label{compare}}
\end{figure*}

The ISM dynamics 
related to the stripping process has significant implications for the X-ray 
luminosity of the galaxy. In fact, we find that the continuous, but 
time-dependent stripping process studied in the present paper provides a 
possible explanation of the observed range of X-ray luminosities of 
large cluster ellipticals. Figure \ref{compare} shows the model luminosities 
of the four cases studied against the $L_B-L_X$ scatter plot obtained from 
X-ray data (Dow \& White 1995; Brown \& Bregman 1998). 

\begin{figure*}[\protect{th!}]
\includegraphics[height=14.cm] {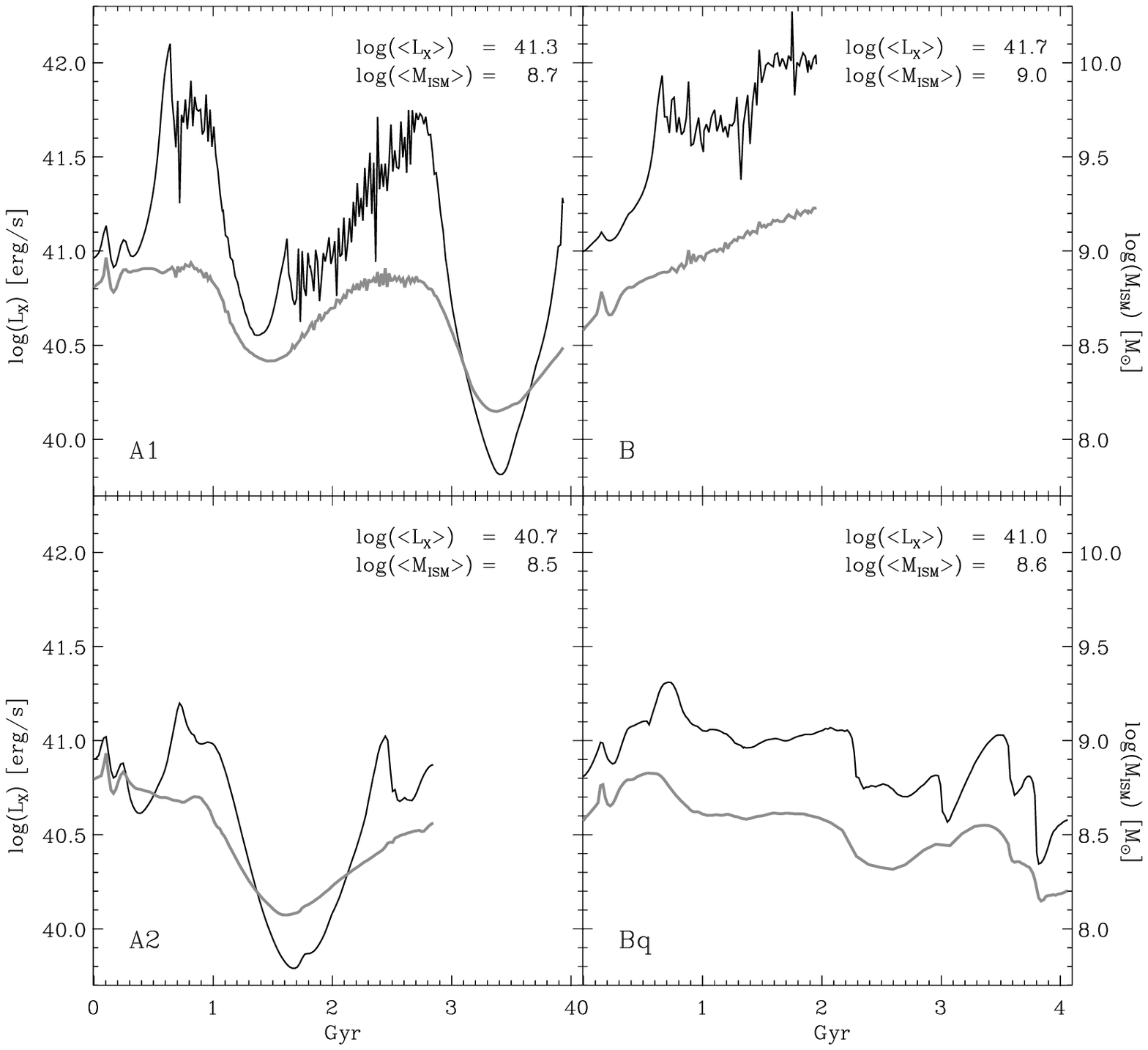}
\caption{\small Gas-mass content and X-ray luminosity of the models as a function 
of time. Black lines refer to the X-ray luminosity in the 0.2-2.4 
keV band; grey lines represent the ISM mass contained within the 
galactic half-light radius $R_l=16$ kpc. Model names are indicated on the 
bottom left of each panel; on the top right of each panel the time averages 
are given. \label{xolm}}
\end{figure*}

The same galaxy may 
display a widely different X-ray luminosity at different times, depending on the 
dynamic `phase' the ISM is involved in (Fig. \ref{xolm}). Just after the 
large pericentric, supersonic (ram-pressure) stripping event, the X-ray 
luminosity is very small; while during the apocentric gas accumulation phase 
it rises to reach a 
maximum in response to the increasing external confinement just preceding a 
new pericentric ram-pressure stripping event. This correlation is evident 
in Fig. \ref{xolm} especially for case A1. Case B, by contrast, is so 
accretion dominated that gas accumulation proceeds unchecked within the 
half-light radius of the galaxy, in spite of substantial stripping in the 
outer layers (see Fig. \ref{machex}). Correspondingly, its X-ray luminosity 
rises continually as the ISM cools and concentrates onto the centre. Ongoing 
stripping mainly results in the large variability, over the scale of one 
crossing time, of the X-ray luminosity.
 
After sufficient time, due to what we have 
called a `secular' evolution of the ISM, most galaxies will end up
either completely stripped, like case A2, or hosting a 
cooling flow, like case B, with correspondingly widely different X-ray 
luminosity. 

\begin{figure*}[\protect{th!}]
\includegraphics[height=14.cm] {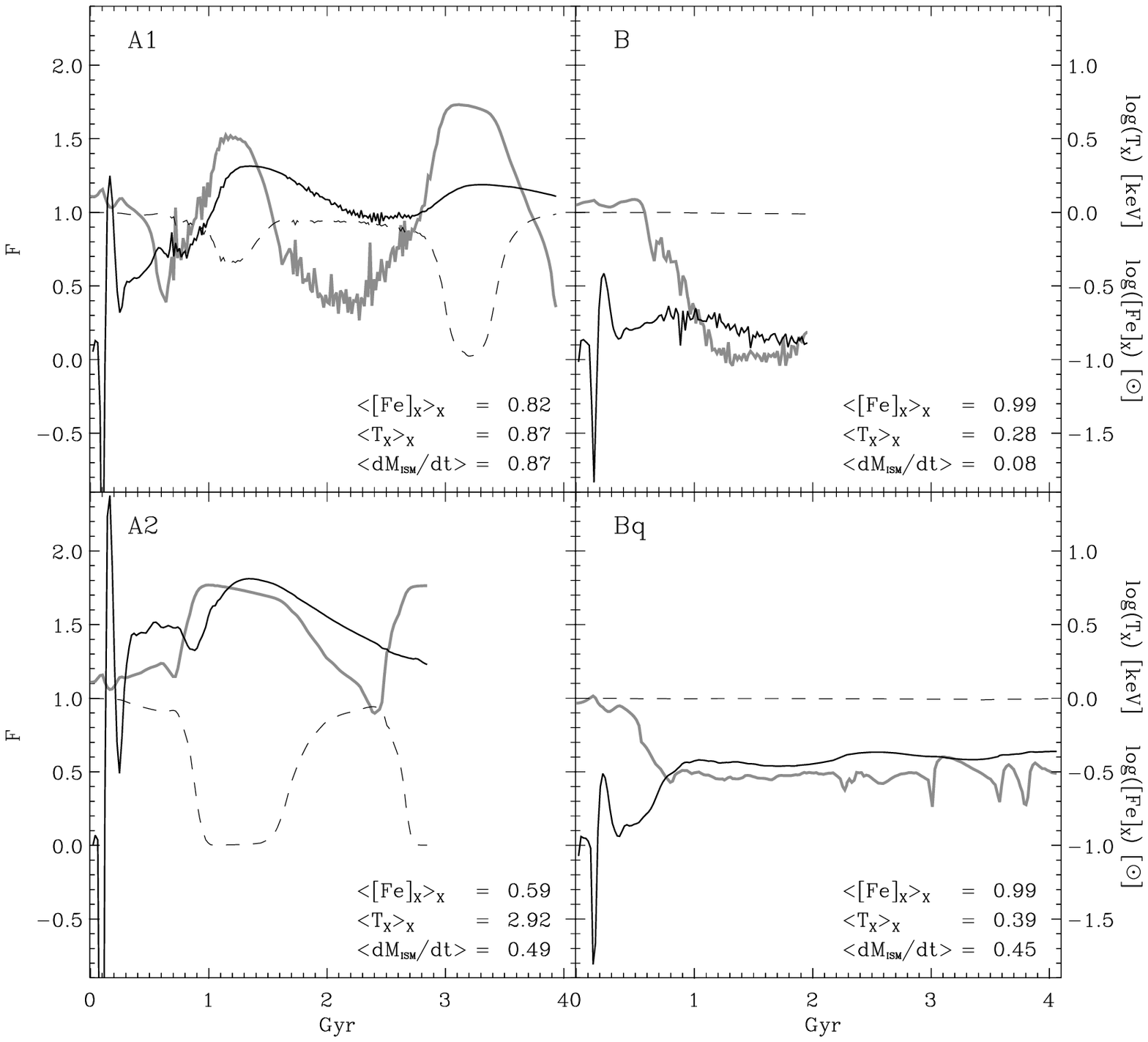}
\caption{\small Stripping fraction and X-ray temperature of the models 
as a function of time. Solid black lines refer to the quantity $F$ defined 
as the ratio of the total gas-mass loss to the total injected gas mass. The 
scale is given on the left side. Thick grey lines represent 
the X-ray emission temperature (scale on the right). The dashed black 
lines are for the iron concentration in solar units. As in Fig. \ref{xolm}, 
the quantities refer to the region of the galaxy comprised within its 
optical half-light radius. On the bottom right of each panel the 
luminosity-weighted time averages of the emission temperature and of the 
iron abundance, as well as the average mass-loss rate are given. 
\label{xlom}}
\end{figure*}

The large scatter in X-ray luminosity between otherwise similar galaxies 
finds a natural interpretation in terms of a time evolution over a timescale 
of the order of the galaxy's orbital period ($\sim$1 Gyr). 
The long-term evolution of the ISM halo (Section 4.4), on the other
hand, which 
brings a galaxy either to be cyclically stripped or otherwise, takes 
several billion years. This time scale is small, but not entirely negligible 
compared to the evolutionary time scale of the cluster itself (Dressler \& 
Shectman 1988).

X-ray temperatures correlate with the fraction of 
stellar ejecta lost from the galaxy. In Fig. \ref{xlom} the quantity 
\be
F=\left[M(0)-M(t)+\int (\dot M_{inj} - \dot M_{do}) dt\right]
\mbox{\Large $/$}\int \dot M_{inj} dt 
\label{fraction}\ee
is shown. In a steady state, $F=1$. If $F>1$, the gas-mass content of 
the galaxy is decreasing, on average; if $F<1$, mass injection prevails 
and ISM is accumulating; $F<0$ indicates that net accretion has occurred. 
After an initial phase, for cases A1 and A2, which end up cyclically stripped, 
$F$ tends to unity. In case B, it tends to zero. Case Bq is intermediate, 
where part of the injected gas is lost, and part is retained, at least over 
a longer time scale. 
Correspondingly, also X-ray temperatures tend to range between extremes in the 
non-drop-out cases, and attain an intermediate value in case Bq. 

Summarizing, model X-ray temperatures do not compare well with the 
data. Typical observed X-ray emission temperatures are about 1 keV. In 
the models, this temperature seems to be avoided. The ISM is either heated 
above the injection temperature in the case of efficient stripping, or it 
cools below it when stripping is inefficient. Case Bq does slightly better 
than the others, but the temperature is still too low, even for a
relatively large value of 
$q$ (see Section \ref{strip-and-drop}). We comment 
on this inconsistency (see also Sect. 6.2) in the Conclusions (Sect. 8). 


\section{X-ray morphology}

\subsection{X-ray emission during supersonic stripping}

\begin{figure}[th!]
\hspace{.4cm}
\includegraphics[width=7.cm]{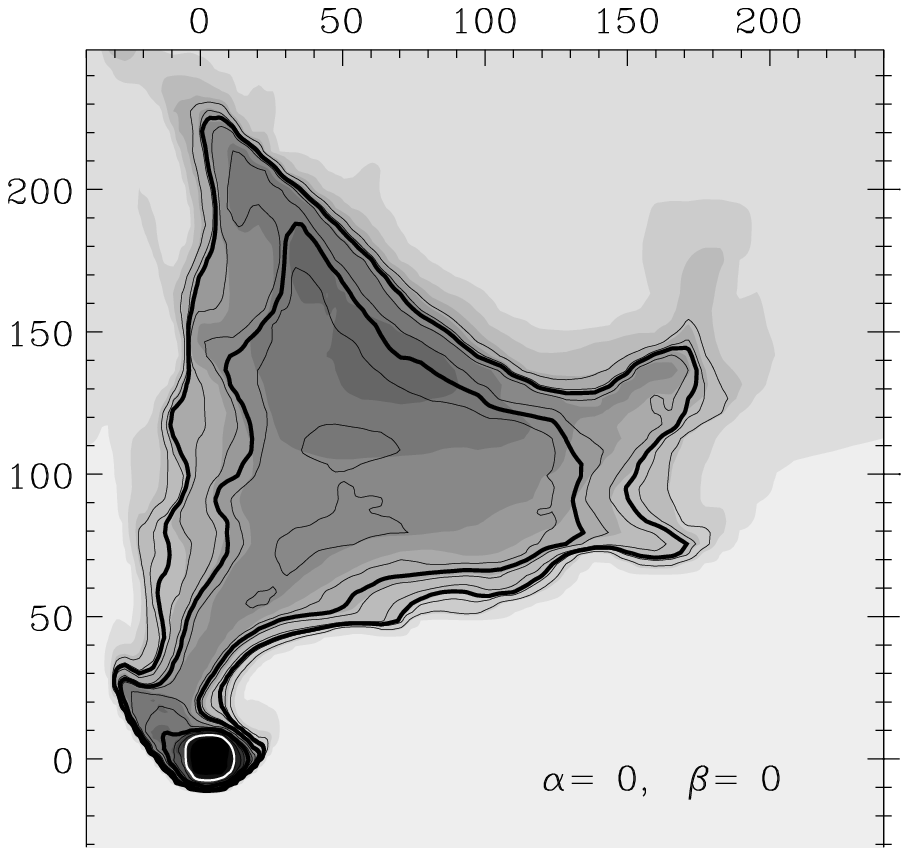}
\raisebox{0.4cm}{\hspace{.5cm}
\includegraphics[width=5.1cm]{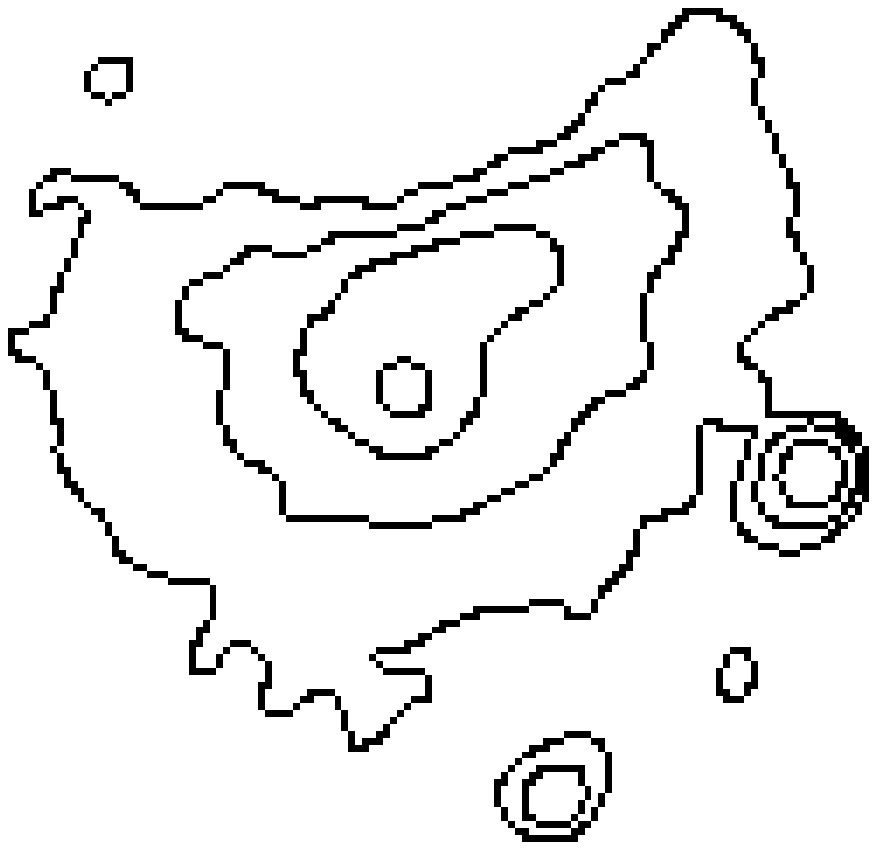}
\hspace{-4.3cm} \raisebox{4.05cm}{
\includegraphics[width=2.4cm]{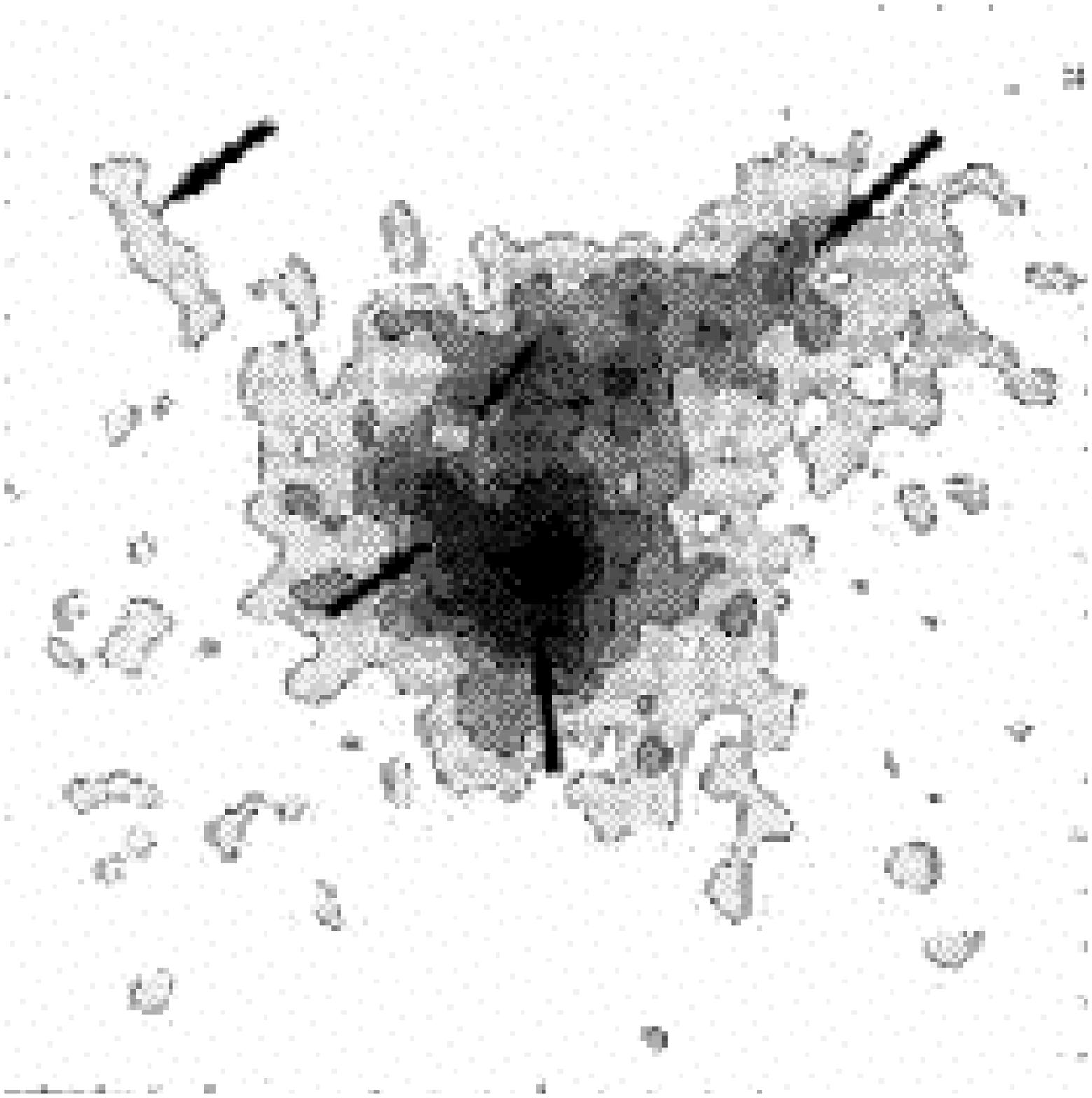}
}}
\vspace{-.7cm}

\noindent
\hspace{.4cm}
\includegraphics[width=14.cm]{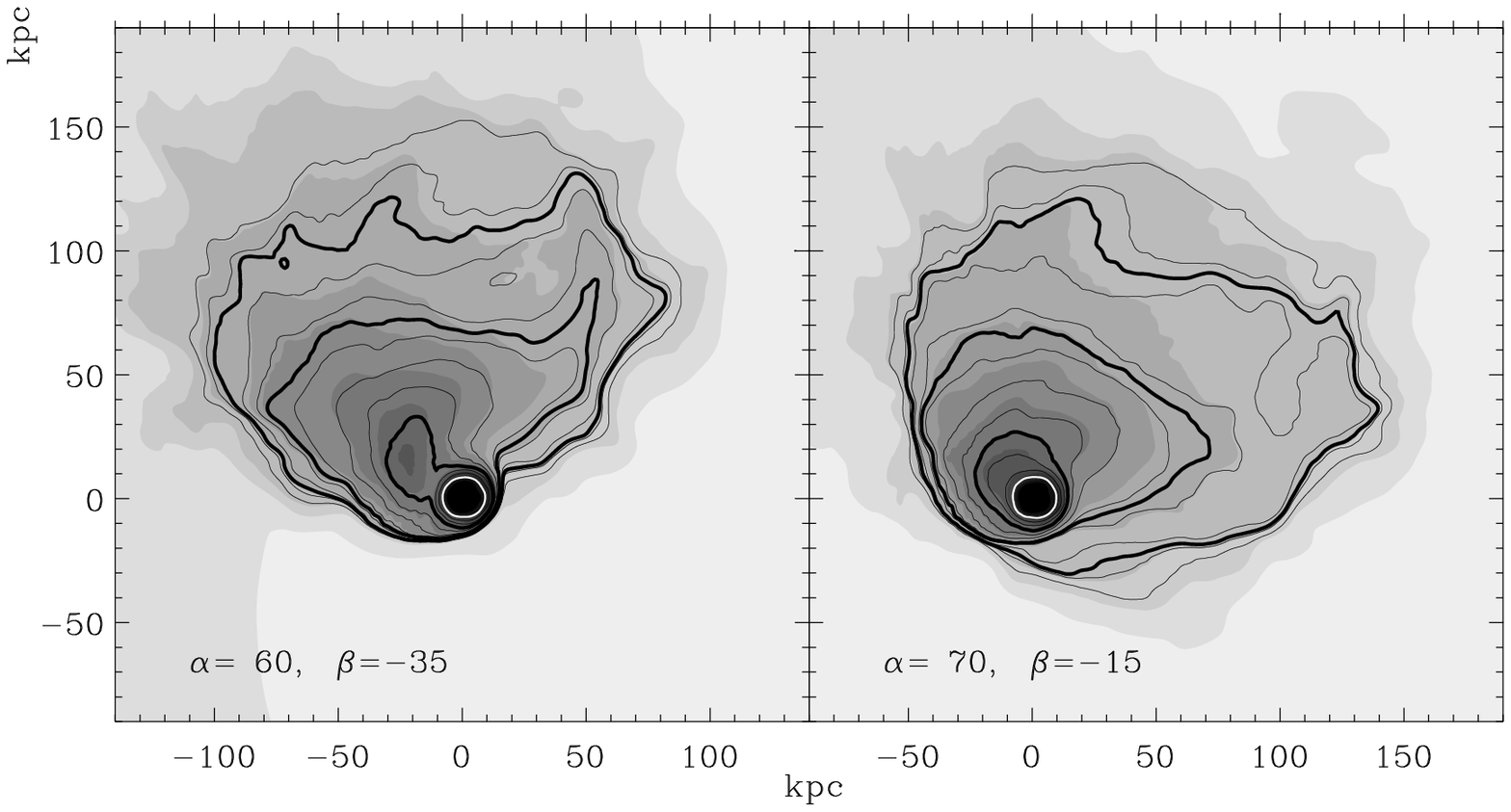}

\noindent
\hspace{.2cm}
\includegraphics[width= 5.cm]{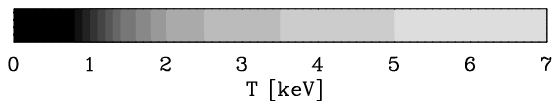}

\caption{\small X-ray maps of model B at $t=579$ Myr in three
different projection directions 
compared to ROSAT PSPC/HRI data for M86 (top right, from Rangarajan et al. 
1995). PSPC data are represented by contour lines, HRI data by 
grey scales. Both observations -- PSPC and HRI -- and the models are
to the same scale, the scale being 
given at the bottom and on the left. The model maps (upper left, and lower 
panels) give the temperature coded in grey scales (scale at the bottom 
of the Figure), and the X-ray surface brightness as contour levels, 
logarithmically spaced from -17.7 to -16.2 in units of erg s$^{-1}$ cm$^{-2}$
arcsec$^{-2}$. The thicker contours are directly comparable to those of the 
PSPC map. The quantities $\alpha$ and $\beta$ given at the bottom of each 
model projection are the polar angles of the projection direction in degrees. 
 \label{xBproj}}
\end{figure}

The stage of ongoing supersonic stripping of the galaxy with its initial 
extended ISM halo lasts between 200 and 400 Myr, depending on the density 
of the ICM. The chances of observing such an event are therefore relatively 
small. However, 
a supersonic stripping process has often been related 
to the X-ray halo surrounding the Virgo elliptical NGC4406/M86
(e.g. Nulsen 1982; Rangarajan et al. 1995). 
This galaxy has a blue shift of 230 km/s, which added to the recession 
velocity of the Virgo cluster of 1150 km/s results in an estimated velocity 
of the galaxy relative to the ICM of 1380 km/s out along the line of sight. 
The galaxy's total X-ray luminosity in the 0.1--2.4 keV band is 
$L_X\sim4\times10^{41}$ erg/s within a radius of 80 kpc, and the emission 
temperature is $T_X\simeq0.8$ keV (Brown \& Bregman 1998). The observed X-ray 
map is shown in Fig. \ref{xBproj} together with some synthetic maps from the 
simulations. Rangarajan et al. (1995) interpret the steep decline in the 
X-ray surface brightness on the north side of the galaxy as a possible Mach 
cone, and from its aperture derive an upper limit to the velocity component 
in the plane of the sky of 700 km/s. Combined with the approaching velocity 
reported above, an estimated streaming velocity of the ICM of $\sim1550$ km/s 
in the frame of the galaxy is obtained. 

The X-ray image of the M86 region 
has probably two major distinctive features which 
are thought to indicate ongoing stripping. One is the sharp decline in 
X-ray surface brightness on the north side, which makes the ROSAT/PSPC 
contour levels almost semi-circular. The other is the emission plume to the 
north-west of the emission maximum, best seen in the ROSAT/HRI data. The 
plume itself is very irregular and shows emission maxima and `holes', as 
well as significant variation in the emission temperature (Rangarajan et al. 
1995). 

The model shown in Fig. \ref{xBproj}, which is case B at $t=579$ Myr, has 
a total X-ray luminosity of $L_X=3.23\times10^{41}$ erg/s, and an 
emission-weighted 
average emission temperature of $T_X=0.73$ keV, comparable to 
those observed for M86. The three projections result in apparent velocities 
in the plane of the sky of 1580, 1126 and 699 km/s, respectively, and 
apparent blue shifts with respect to the cluster of 0, 1109 and 1417 km/s, 
respectively. Thus especially the last projection may be suitable for a 
comparison. The synthetic X-ray surface-brightness distribution has an overall 
shape similar to that of the M86 region. 
However, neither `plume' nor `edge' are visible 
in this projection. Intrinsically, both features are present in the model, as 
is seen from the projection onto the equatorial plane. We are unable, though, 
to find a viewing direction from which both can be distinguished. The 
plume is quite prominent if the velocity in the plane of the sky is large; 
the rear edge of the ISM distribution, however, is evident only for very 
small relative blue shifts, and then the overall shape of the contours is 
triangular and not semicircular. When comparing the model maps to the data 
for M86,  the coincidence of the X-ray centroid and the X-ray maximum in the 
latter becomes particularly puzzling. If the models under discussion are truly 
relevant for M86, we find that the Mach cone interpretation of Rangarajan et 
al. (1995) should be considered with some caution, allowing also for different 
possibilities. For instance, the edge may be the result of a strong 
inhomogeneity in the surrounding ICM. Also, the observed X-ray halo is 
probably related to a whole subcluster of $\sim$50 members rather than to 
M86 alone, resulting in a wider, possibly more irregular, potential
well (Binggeli et al. 1993; Schindler et al. 1999).
 
A comparison with our models is still significant, though, because the 
gravitational potential and the gas sources near M86 should still be dominated 
by the giant elliptical. 

Figure \ref{xBproj} gives some indication as to whether 
a detailed X-ray temperature map of 
an object like M86 would be helpful to determine its instantaneous motion 
relative to the ICM. We find that the synthetic X-ray temperature maps 
generally follow the X-ray surface-brightness maps, higher emissivities 
corresponding to colder gas. However, some remarkable deviation could be 
measured by an instrument sensitive to variation of 0.1 keV in temperature 
at $T\simeq1$ keV. 
Specifically, the `plume' in the $\alpha=60$, $\beta=-35$ projection is 
colder than the `surrounding' gas, in projection. By comparison with the 
$\alpha=0$, $\beta=0$ map, it is seen that the gas visible in the plume is 
actually quite far away from the centre of the galaxy, and is cooler not 
as a result of radiative cooling but of the supersonic expansion of the 
stripped ISM downstream of the galaxy. Thus, an interpretation of the plume 
as of a `blob' of freshly stripped gas, as opposed to that of the northern 
edge, would be misleading. From a single projection the kinematics of the ISM 
are not unambiguously identifiable. 

%
In general, from our models we find that 
there is not necessarily an equivalence between the gradient in X-ray 
surface brightness, the pressure gradient, and the amount of ram pressure. 
%
The modelled ISM 
is not in equilibrium with the ambient pressure, but it expands (and 
even cools) while it is removed from the galaxy. 
Steep gradients in surface brightness partly arise from projection effects.
Conversely, if 
the model is viewed in direction of the motion of the galaxy while
it is subject to stripping, there is no obvious 
indication that stripping is 
occurring. From the synthetic X-ray
maps one can work with the hypothesis that the gas is in hydrostatic 
equilibrium without meeting any evident inconsistencies, except for
grossly misestimating the gravitational field 
of the galaxy. 
%


\subsection{X-ray surface-brightness profiles}

\begin{figure}[b!]
\begin{center}
\includegraphics[width=14.cm]{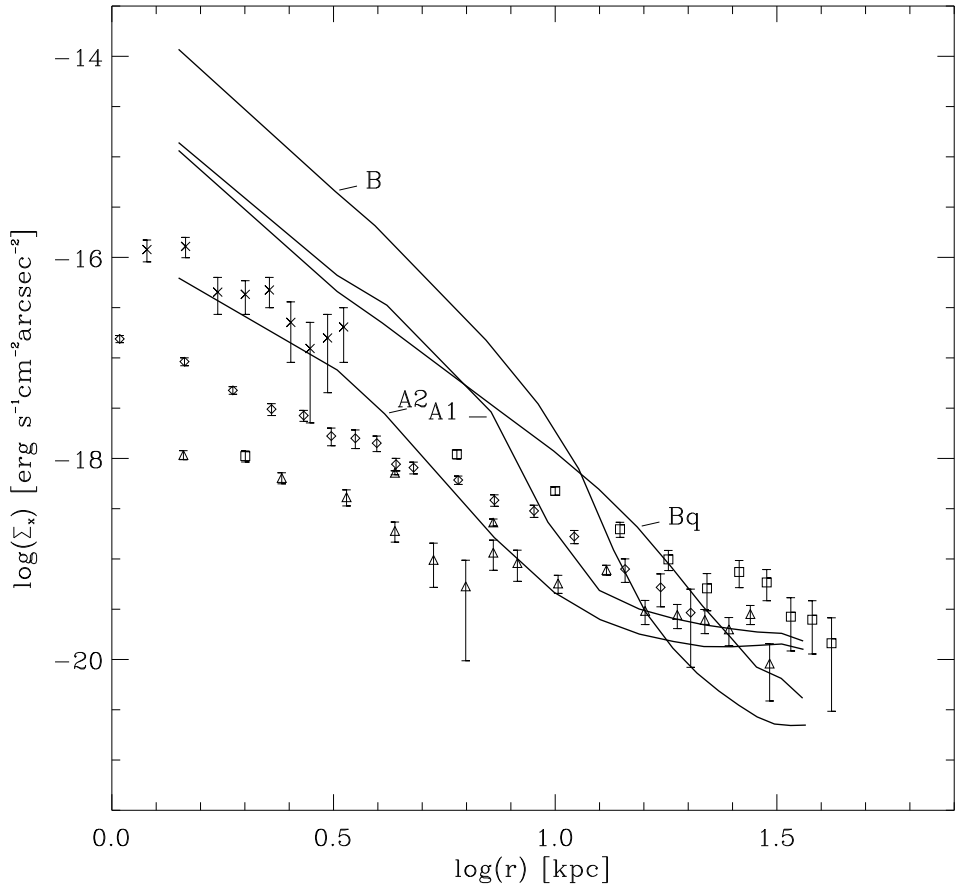}
\end{center}
\caption{\small Examples of azimuthally averaged 
radial X-ray emission profiles as 
derived from the models, compared with X-ray data of the elliptical 
galaxies NGC 4374 (triangles, Forman et al. 1985), NGC 4649 (diamonds, 
Trinchieri et al. 1997), and NGC 1404 (squares, Fabbiano et al. 1992), 
and of the lenticular galaxy NGC 1380 (crosses, Schlegel et al. 1998). 
The curves refer to the cases `A1', `A2', `B', and `Bq' as labelled, in 
projection onto the $x-z$ plane (except for case `B' which is
projected onto 
the $y-z$ plane), and at times $t= 1972$, 1973, 1862 and 1868 Myr, 
respectively. 
\label{xprof}}
\end{figure}

If the majority of cluster galaxies are subject to stripping, an important 
test of the present model is the comparison of typical surface brightness 
profiles of cluster galaxies with those obtained from the simulations. 
After a time of the order of the radial orbital period, when the outer ISM 
halo is subject to continuous stripping, the ISM distribution tends to become 
quite compact as a result of pressure confinement, continuous loss of gas 
from  regions where the replenishment rate is small, and cooling. Viewed from 
a direction within $\pi/4$ of that of the velocity of the galaxy, the 
contours of constant X-ray surface brightness are fairly circular. 
Examples of azimuthally averaged emission profiles obtained for such cases are 
shown in Fig. \ref{xprof} for all four models, and compared to the X-ray 
data of NGC 4374, NGC 4636, NGC 1404 and NGC 1380. The first two of these 
galaxies are Virgo cluster members, while the latter two are in the
Fornax group  
dominated by NGC 1399. They represent cases in which ISM mass loss
caused by  
gas dynamic interaction with the environment may have occurred. 

Fig. 14 shows that the model profiles do not compare well with 
X-ray data. They are too steep, with most of the emission originating in the 
central 10 kpc. 
The 
inclusion of substantial drop out in case `Bq', although making the 
emission more regular and more extended, does not significantly improve the 
fit to the data. Model `A2'  matches rather better at 
least the data of NGC 1380; the model total luminosity and the 
average emission temperature have values comparable to those of NGC 1380 
($L_X=3\times10^{40}$ and $T_X\simeq0.5$ keV). This result, though, is not 
robust, in that it depends on numerical implementation details and on the 
particular time chosen. Essentially, it appears that cooling of 
the gas in the centre of the galaxy is too efficient in the models. This 
is shown by computations with four times higher resolution in each spatial 
dimension performed for cases `B' and `A2'. 
In spite of the large angular momentum of the ISM 
which tends to inhibit spherical central accretion, the profiles computed on 
the high-resolution grid are even more `bimodal', with a distinct core
of dense  
ISM. 
%

In conclusion, the present models suffer from the same problem as 
`homogeneous' steady-state cooling-flow models, namely the excess 
in central emission. This shortcoming is even increased by the 
stripping truncation of the ISM halo, which makes the emission profiles 
steeper overall, in such way that even the $q$-parameter drop-out 
prescription does not seem to be helpful. 

\section{Diluted ISM in the Cluster}


%
\subsection{Iron Enrichment and Distribution}

The stripping of cluster galaxies causes the iron synthesized in SNIa events to 
mix with the ICM. Adopting the average stripping rate of our models after the 
first stripping, $\sim1\msolar$/yr, we derive a net iron injection rate into the 
ICM of $\sim0.2\msolar$/yr, if we assume a solar abundance in the ISM and a central 
density of 300 galaxies per Mpc$^3$ within a core radius of $\sim500$ kpc, like 
in Coma (Biviano et al. 1995). This figure doubles if $\sim10$\% 
of the galaxies in the cluster core are experiencing the stripping of an extended 
tenuous halo, like our models initially. 

In young clusters, the ICM enrichment is even larger, because young stellar 
systems shed more gas to the environment (Eq. \ref{Renzini}). 
In a computation for such case (model ``C2'' of Toniazzo 1998), 
in which the thermonuclear age of the galaxy was taken to be $\tau_*=3.44$ Gyr, 
the total amount of iron injected 
into the ICM over 2 Gyr was $\sim6$ times as large as that of model A. Integrating 
the iron shed over $\tau_*=$1--15 Gyr under a `steady-stripping' assumption, we 
derive a total mass of iron in the core of a Coma-like cluster of 
$1.3\times10^{10}\msolar$ from galaxy stripping. For comparison, using the total 
gas mass and the gas density profile given by Reiprich (1998), and the metallicity 
given by Fukazawa et al. (1998) in the Coma cluster, the total amount of iron in 
the core ($R<500$kpc) is estimated to be $2.6\times10^{10}\msolar$. 
Thus the standard scenario of stellar evolution, combined 
with the steady stripping of cluster galaxies plus a $\sim$10\% 
of infalling objects, can explain the iron content of the ICM in a Coma-like 
cluster. 

\begin{figure}[h!]
\begin{center}
\includegraphics[width=15.5cm]{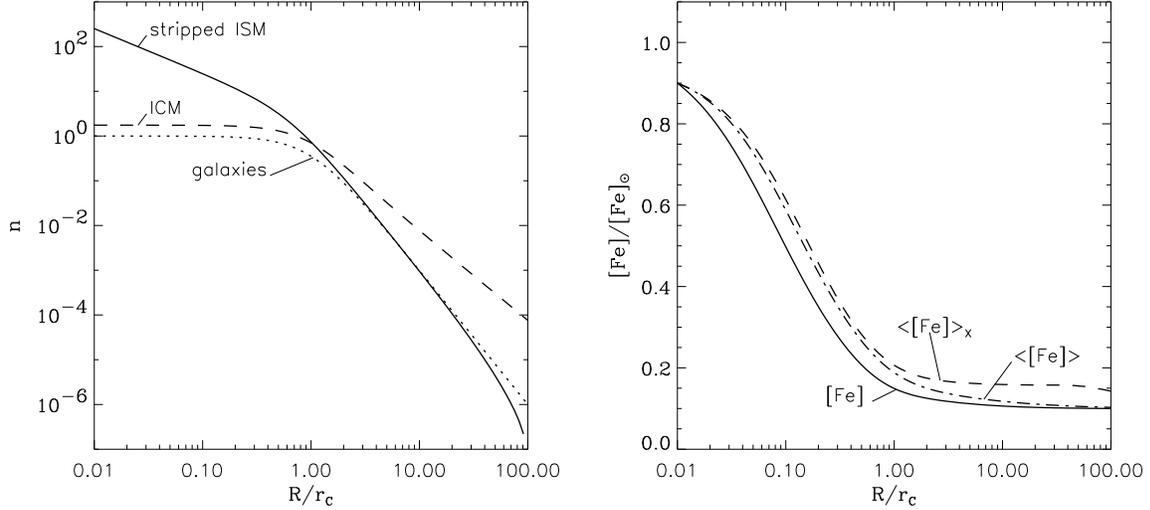}
\end{center}
\caption{\small Left: Number density of cluster galaxies normalized to 
the central number density (dotted line), the corresponding normalized 
pericentral density of stripped ISM (solid line), and the 
ICM density for the cluster mass distribution defined by Eqs. 
(\ref{cluster_pot}) and (\ref{rhoC}), normalized to the same total gas mass. 
The radial coordinate is in units of the cluster core radius $r_C$. 
A mixture of 95\% 
``ICM'' gas distribution and of 5\%
``stripped ISM'' gas distribution is assumed. 
Right: abundance of iron in the cluster gas as obtained from Eq. 
(\ref{Feprofile}) (solid line -- see text), mass-weighted average iron
abundance  
of the intracluster gas within a given radial distance (dash-dotted line), and 
emission-weighted average iron abundance of the ICM (dashed line). 
\label{perdist}}
\end{figure}

The injection rate of cool ISM into the ICM follows the number density of 
the cluster galaxies weighted by their stripping rate. Stripping mainly 
occurs while the galaxy is passing its orbital pericentre. 
Therefore, as opposed to a steady stripping process in which the 
ISM-shedding rate is constant and the iron enrichment follows the spatial 
distribution of galaxies, one can consider the case in which each galaxy 
deposits all gas stripped during one radial orbital period at the location 
of the orbital pericentre. The resulting radial density distribution of enriched 
material is proportional to the volume number density of 
cluster galaxies as a function of their pericentral distance, $n_{gp}(R_p)$, 
defined in such a way that $n_{gp}(R_p)4\pi R_p^2dR_p$ is the number of cluster 
galaxies which follow an orbit with a pericentre comprised between $R_p$ and 
$R_p+dR_p$. If the velocity distribution of the cluster galaxies is assumed to 
be isotropic and isothermal, and their spatial density is given by the function 
$n_g(R)$, where $R$ is the radial distance from the cluster centre, 
the function $n_{gp}(R_p)$ is given by 
\be
n_{gp} = \int_1^\infty\left(1+z_f-z_{rp}\right){\rm e}^{-z_f/2y^2}
   \frac{yn_g(yx_p)}{\sqrt{y^2-1}}dy \, , \label{peridist} \ee
where $x_p\equiv R_p/r_C$, $r_C$ is the cluster scale radius, 
$z_f=2y^2[\phi(yx_p)-\phi(x_p)]/(y^2-1)$, 
$z_{rp}=x_pd\phi(x_p)/x_p$, and $\phi\equiv\Psi_C/\sigma_C^2$ is the cluster
gravitational potential normalized to the (constant) one-dimensional velocity 
dispersion. The function $n_{gp}$ is shown in Fig. \ref{perdist} for
the example 
of the cluster mass distribution and cluster gravitational potential of 
Eq. (\ref{cluster_pot}), with $n_g(R)=1/(1+x^2)^{3/2}$ and $x\equiv R/r_C$. 
The function $n_{gp}$ is steeper than $n_g$ and for $x<1$ it is $\sim1/x$. 
These properties are very general and do not depend on the particular choice 
of the cluster density profile.

If the main source of iron enrichment of the ICM is actually due to the stripping 
of cluster galaxies, the function 
\be {\rm [Fe]} = {\cal F}(R) = 
\frac{{\rm [Fe]}_{ICM}(1-S)\rho_C/\rho_{C0} +  [Fe]_{ISM}Sn_{gp}(R/r_C)/n_g(0)}
     {(1-S)\rho_C/\rho_{C0} +  Sn_{gp}(R/r_C)/n_g(0)}
 \, , \label{Feprofile}\ee 
with the `unperturbed' ICM density $\rho_C$ given e.g. by Eq. (\ref{rhoC}), 
will be proportional to the [Fe] abundance in the cluster gas. Here, 
$S$ is the ratio of the total gas mass of stripped material present in the ICM 
to the total ICM gas mass. The function 
${\cal F}(R)$ is shown in Fig. \ref{perdist} for the cluster model used 
for the calculations of model `A1' and `A2', and the parameters [Fe]$_{ICM}$=0.1, 
[Fe]$_{ISM}$=1, and $S=0.05$. In the same plot also the profiles are shown for 
the total mass fraction of iron within a given radius as a function of radius, 
and for the corresponding total X-ray emission weighted iron abundance.
The occurrence of negative cluster 
metallicity gradients such as that shown in Fig. 15 is confirmed  
by recent X-ray observations as spatially resolved spectroscopic 
X-ray data become increasingly available (e.g. Molendi \& De Grandi 1999).

\subsection{Infall into the Cluster Centre}

The inhomogeneous enrichment of the ICM by stripped material 
can have a further impact on the abundance distribution and 
also on the temperature of the cluster gas in the centre.
With typical density ratios of 
100-1000, the ISM lost from cluster galaxies will tend to fall inwards. 
We can argue that the mass flux involved in a `galaxy stripping
induced inflow', in which the diluted ISM resulting from cluster 
galaxy stripping falls into the centre of the cluster, is a
monotonically increasing function of radial distance. 
The reason lies in the limited survival time of the stripped clouds. 

The mass flux caused by the infall of stripped ISM clouds is 
\be \dot M = 4\pi R^2 n_{gal} \Delta R \dot M_{st} \, , \label{stripin} \ee
where $R$ is the radial location 
in the cluster, $\dot M_{st}$ the mean gas mass stripping rate for a galaxy, 
and $\Delta R$ the average distance traveled by a cloud. 
The cloud is accelerated radially by $-(1-1/\tilde r)d\Psi_C/dR$ for a time 
$\ds t_s=F\frac{D}{\sigma_C}\sqrt{\tilde r}$, with $D$ the linear size of 
the cloud, $\tilde r=\rho_{ISM}/\rho_{ICM}$ and  $F$ a factor of order unity. 
After a time $t_s$ the cloud is destroyed by the development of Rayleigh-Taylor 
instabilities (see e.g. Nittmann et al. 1982). Note that $\sigma_C$ is both the 
typical initial velocity of the cloud, and the sound speed of the hot ambient 
medium. 
The distance traveled by the cloud during the time $t_s$ is thus 
$\ds \Delta R \simeq F^2D^2 \left(\tilde r-1\right) 
\frac{R/r_C^2}{1+(R/r_C)^2}$.

If we use the distribution of Eq. (\ref{peridist}) for the cluster galaxies, 
by interpreting  
$R$ as the pericentre of the galaxy orbit, the resulting mass-flux profile, 
Eq. (\ref{stripin}), is 
$\ds \dot M(R) \simeq 4\pi F^2D^2(\tilde r-1)\dot M_{st}r_C x^2n_{gp}(x)/(1+x^2)$, 
which is $\sim R^2$ in the 
centre, has a maximum near $R\simeq 0.8R_C$, and decreases as $1/R^2$ for $R>1$. 
The mass flux variation, however, is not associated with 
drop out but with actual accumulation of hot gas in the cluster centre. The radial 
divergence of $\dot m_{st}/R^2$ is proportional to $n_{gp}\sim1/R$ 
for $R<r_C$. The ICM thus tends to accumulate in the cluster centre on a 
$1/R$ profile. 

With $F\simeq3$ (Nittmann et al. 1982), $r\simeq10$,  $D/r_C\simeq0.01$, 
$\dot M_{st}\simeq1\msolar/\yr$, and a total number of 
$\sim10^3$ cluster galaxies within the cluster
core radius, a maximum non-cooling inflow of $\sim15\msolar$/yr is found at 
$R=0.79 r_C$. Note that this value depends on the number of galaxies within the 
cluster core, and on the mean stripped ISM cloud size as well. The latter 
decreases with $\dot M_{st}$, since the size of the vortices 
determining the size of the stripped clouds is also smaller. Thus, a 
significant inflow rate occurs only in clusters in which large 
stripping events take place.

\section{Summary and Concluding Remarks}

We have studied the gas dynamic stripping of an elliptical galaxy orbiting in 
a cluster of galaxies for four cases (or ``models'') by use of time dependent, 
three-dimensional, Eulerian, hydrodynamic numerical simulations. Accurate 
modeling of the galaxy, of its orbit, of the cluster background, and of the gas 
dynamics results in significantly reduced ISM mass-loss rates compared to previous 
calculations. 

An initial state was chosen to represent a gas-filled galaxy which is entering the 
cluster on a rosette-shape orbit. The computation was continued until an 
unambiguous long-term evolution of the ISM halo was established for each case. 

The gas dynamic evolution is different in different sections of the galaxy's 
orbit. Near the pericentre, when the velocity of the galaxy is greater than the 
speed of sound in the ambient medium, ram pressure is effective. A major, 
sudden stripping `event' can occur if the kinetic pressure grows rapidly to a value 
of the order of the central thermal pressure in the galaxy ($\sim0.02$ keV 
cm$^{-3}$ for our models), and over a time comparable to or smaller than the 
sound crossing time of the ISM halo ($\sim50$ Myr). Otherwise, stripping is more 
gradual. 

Near the apocentre, where the orbital speed is slightly smaller than the speed of 
sound, the evolution is governed by gas-mass replenishment, radiative cooling, and 
the growth of Kelvin-Helmholtz modes. The latter result in the formation of 
eddies, that entrain streaming as well as galactic gas, and are eventually dragged 
away. This gives ``impulsive'' stripping alternated with short accretion phases. 
During this phase, mixing of the ISM and the ICM within the galaxy is significant. 

Of the four cases studied, two (``A1'' and ``A2'') evolve, with different speeds, 
towards a cyclic ``stripping replenishment'' dynamics, in which stripping near the 
orbit's pericentre is nearly complete, while a small ISM halo re-forms in the 
``upper'' half of the orbit. In one case (``B''), cooling proceeds almost 
undisturbed and a ``cooling flow'' develops within the galaxy's
half-light radius,  
with the cooler gas distributed in a large disk of $\sim$20 kpc diameter. The 
angular momentum of the ISM is partly generated by ram-pressure torque in the 
supersonic phase, and partly by accreted orbital angular momentum in the subsonic
phase. 

The efficiency of the stripping process was estimated in terms of a dimensionless 
parameter, $\eta$, which gives the magnitude of the mass loss over a radial orbital 
period when linearly combined with cooling and replenishment to describe the 
long-term evolution of the gas-mass content of the galaxy. It appears 
that $\eta$ depends mainly on the orbit of the galaxy and on the cluster 
environment, and to a lesser extent on the total mass or the total optical 
luminosity of the galaxy. 

The fourth model (``Bq'') was identical to model ``B'' except for the inclusion 
of mass drop out in the gas dynamic computations. The drop-out prescription 
adopted follows Sarazin \& Ashe (1989) with $\dot\rho_{do}=q\rho/\tau_c$, 
$\tau_c$ being the cooling time and $q$ a dimensionless factor, which was taken 
to be $q=0.4$. It was found that drop out heavily affects the gas dynamics, causing 
enhanced total gas mass loss from the galaxy. Model ``Bq'' eventually approaches 
the cyclic stripping dynamics of models ``A1'' and ``A2'', if on a much longer 
time scale. 

The computed models account fully both for the observed correlation between 
optical and X-ray luminosities of cluster elliptical galaxies, and for its 
large dispersion. The X-ray luminosity of the gas associated with the model 
galaxies varies strongly depending on its orbit and on time. 
This time variation can produce for the same optical galaxy the whole range 
of X-ray luminosities observed. Due to the large gas dynamic energy input, 
stripping reduces the average X-ray luminosity $L_X$ of the ISM halo much less than 
its average mass. In this case drop out mainly reduces the 
range of $L_X$. 

The X-ray morphology, luminosity and temperature of the models during the 
first, large stripping phase after ``infall'' are consistent with the 
X-ray data of the region surrounding the Virgo cluster 
elliptical M86, where gas dynamic stripping of an extended gaseous halo 
is taking place (Rangarajan et al. 1995). 
However, it was also found that after the ISM halo has suffered substantial 
truncation due to stripping, the computed X-ray temperatures and X-ray morphologies 
of the models are inconsistent with X-ray data of X-ray fainter cluster 
ellipticals, being generally too cool and too centrally concentrated. The 
inclusion of drop out does not improve the situation significantly. 

We think that we are encountering here a difficulty inherent to the 
hydrodynamic treatment adopted, common also to steady-state, so-called 
homogeneous cooling flow models (e.g. Sarazin \& White 1987), which
have too steep X-ray profiles as compared to the data. 
It is known that the ISM is not thermally stable (Kritsuk 1992). 
In a quasi-hydrostatic, steady-state case, 
self-regulation can be achieved via the inclusion of a drop-out term (Kritsuk 
1995). The $q$-prescription allows steady-state cooling flow models to match 
the observed X-ray surface brightness profiles (Bertin \& Toniazzo 1995). 
The stripping process, however, results
in a further 
concentration of the X-ray emission from the ISM, causing more intense 
cooling in the centre and heating in the outer part, thus destroying the 
approximate balance between radiative emission and SNIa heating of 
steady-state models. An even increased value of the parameter $q$ may 
partially solve the problem, at the price, however, of making stripping very 
effective, and thus of greatly reducing the gas-mass content and X-ray 
luminosity of a galaxy, to an extent which seems to be in disagreement 
with X-ray 
evidence on cluster galaxies. 

Table \ref{tabsum} summarizes our main numerical results.

\begin{table}
\begin{center}
{
\renewcommand{\arraystretch}{1.6}
\begin{tabular*}{13.cm}{@{\extracolsep{\fill}}l@{\hspace{3cm}} c c c c}
\hline
Model      &   A1  &   A2  &   B   &   B$q$  \\
\hline
 $\ds M_T${\small\raisebox{2mm}{$^{~~(1)}$}} &  11  &  5.7  &  10  &  10  \\
 $\ds L_B${\small\raisebox{2mm}{$^{~~(2)}$}} &  11  &  5.8  &  11  &  11  \\
 $\ds \left(M_g\right)_0${\small\raisebox{2mm}{$^{~~(3)}$}} 
                                             &  24.2 & 20.7 & 25.6 & 25.6 \\
 $\ds \tau_{c,0}${\small\raisebox{2mm}{$^{~~(4)}$}} 
                                             & 1.7 & 2.0 & 1.6 & 1.6 \\
 $\ds \tau_{orb}${\small\raisebox{2mm}{$^{~~(5)}$}}  
                                             & 1.9 & 1.9 & 1.5 & 1.5 \\
 $\ds \bar m${\small\raisebox{2mm}{$^{~~(6)}$}}   
                                             &  1.23 &  1.23  &  0.95  & 0.95 \\
 $\ds \bar\sigma${\small\raisebox{2mm}{$^{~~(7)}$}}  
                                             & 1.31 & 1.31 & 0.15 & 0.15 \\
 $\ds q${\small\raisebox{2mm}{$^{~~(8)}$}}   &  0  & 0  & 0  & 0.4 \\
\hline
 $\ds \eta${\small\raisebox{2mm}{$^{~~(9)}$}} 
                                             & 4.2 & 4.3 & 2.6 & 3.7 \\
 $\ds F${\small\raisebox{2mm}{$^{~~(10)}$}}  
                                             & 1.1 & 1.2 & 0.1 & 0.6\\
 $\ds \langle L_X \rangle${\small\raisebox{2mm}{$^{~~(11)}$}}  
                                             & 19.4 & 5.3 & 52.7 & 8.9 \\
 $\ds \left(L_X\right)_{max}${\small\raisebox{2mm}{$^{~~(12)}$}}  
                                             & 126 & 15.8 & 188 & 20.4 \\
 $\ds \left(L_X\right)_{min}${\small\raisebox{2mm}{$^{~~(13)}$}}  
                                             & 0.6 & 0.6  & 10 & 2.2 \\
 $\ds \langle T_X \rangle${\small\raisebox{2mm}{$^{~~(14)}$}}  
                                             & 1.53 & 3.13 & 0.51 & 0.38 \\
 $\ds \left(T_X\right)_{max}${\small\raisebox{2mm}{$^{~~(15)}$}}  
                                             & 5.39 & 5.87 & 1.22 & 1.03 \\
 $\ds \left(T_X\right)_{min}${\small\raisebox{2mm}{$^{~~(16)}$}}  
                                             & 0.18 & 0.79 & 0.10 & 0.18 \\
 $\ds \langle [{\rm Fe}]_X \rangle${\small\raisebox{2mm}{$^{~~(17)}$}}  
                                             & 0.72 & 0.52 & 0.99 & 0.98 \\
\hline
\end{tabular*}\\ }
\end{center}
\caption{\label{tabsum} \small \small 
Main numerical results.
\null$^{\,(1)}$: total gravitational mass of the galaxy in units of 
$10^{11}\msolar$.\hspace{.5cm} 
\null$^{\,(2)}$: optical blue luminosity of the galaxy in units of 
$10^{10}\lsolar$.\hspace{.5cm} 
\null$^{\,(3)}$: total mass of the initial ISM gas distribution.\hspace{.5cm} 
\null$^{\,(4)}$: central cooling time of the initial ISM gas
distribution in Gyr..
\hspace{.5cm} 
\null$^{\,(5)}$: radial orbital period in Gyr.\hspace{.5cm} 
\null$^{\,(6)}$: time-averaged Mach number of the galaxy velocity relative to 
the ICM.\hspace{.5cm} 
\null$^{\,(7)}$: time-averaged incoming momentum flux of the ICM in the frame 
of the moving galaxy in units of $10^{-11}$ g cm$^{-2}$ s$^{-1}$.\hspace{.5cm} 
\null$^{\,(8)}$: gas mass drop-out parameter.\hspace{.5cm} 
\null$^{\,(9)}$: gas dynamic stripping efficiency parameter (Equation 
\ref{stimasec}).\hspace{.5cm} 
\null$^{\,(10)}$: Fractional stripping (Eq. \ref{fraction}).\hspace{.5cm} 
\null$^{\,(11)}$: average ISM X-ray luminosity in units of $10^{40}$ erg/s.
\hspace{.5cm} 
\null$^{\,(12)}$: maximum ISM X-ray luminosity in units of $10^{40}$ erg/s.
\hspace{.5cm} 
\null$^{\,(13)}$: minimum ISM X-ray luminosity in units of $10^{40}$ erg/s.
\hspace{.5cm} 
\null$^{\,(14)}$: average ISM X-ray emission temperature in keV.\hspace{.5cm} 
\null$^{\,(15)}$: maximum ISM X-ray emission temperature in keV.\hspace{.5cm} 
\null$^{\,(16)}$: minimum ISM X-ray emission temperature in keV.\hspace{.5cm} 
\null$^{\,(17)}$: average ISM emission-weighted iron abundance in solar 
units.\hspace{.5cm} 
\label{tabres}}
\end{table}

\vspace{1.5cm}
\begin{center} {\bf Acknowledgements} \end{center}
This work is based on the results of the PhD Thesis of T. Toniazzo, carried 
out at the Max-Planck-Institut f\"ur Extraterrestrische Physik, Garching. 
T. Toniazzo wishes to express his gratitude to R. Treumann, whose supervision 
was extremely helpful, and to G. Bertin, for his continued support. 

Many thanks to H. B\"ohringer and G. Morfill for their support, to 
E. M\"uller for providing a version of PROMETHEUS, to T.W.Hartquist for 
useful comments and to A. Shukurov for stimulating discussions. We
thank the anonymous referee and Phil James for improving the
presentation of the paper considerably.

This work was made possible by the excellent technical support provided 
by the Rechenzentrum of the Institut f\"ur Plasmaphysik, Garching, and by 
financial support from the Scuola Normale Superiore di Pisa, University of 
Trieste, and Deutscher Akademischer Austauschdienst. 

\newpage

\begin{center} {\large \bf References} \end{center}

\begin{itemize}

\item[]
Abadi, M.G., Moore, B., \& Bower, R.G., 1999,  {\it Monthly Not. of the Royal 
Astron. Soc.} 308, 947

\item[]
Anders, E., \& Grevesse, N., 1989, {\it Geochim. Cosmochim Acta} 53, 197

\item[] Arnaud, M., Aghanim, N., Gastaud, R., Neumann, D.M., Lumb, D.,
et al., 
2000, astro-ph/0011086

\item[]
Balsara, D., Livio, M., \& O'Dea, C.P. 1994, {\it Astrophys. J.} 437, 83

\item[]
Bertin, G., Saglia, R.P., \& Stiavelli, M., 1992, {\it Astrophys. J.} 384,423

\item[] 
Bertin, G., \& Toniazzo, T., 1995, {\it Astrophys. J.} 451, 111


\item[]
Binggeli, B., Popescu, C.C., \& Tammann, G.A., 1993 {\it Astron. \& 
Astrophys. Supp.} 98, 275

\item[]
Binney, J., \& Tremaine, S., 1987, {\it ``Galactic dynamics''}, Princeton 
University Press, Princeton 1987

\item[]
Biviano, A., Durret, F., Gerbal, D., Le Fevre, O., Lobo, C., Mazure, A., \& 
Slezak, E., 1995, {\it Astron. \& Astrophys. Suppl.} 111, 265

\item[]
Briel, U.G., Henry, J.P., \& B\"ohringer, H., 1992, 
{\it Astron. \& Astrophys.} 259, 31

\item[]
Brown, B.A., \& Bregman, J.N., 1998, {\it Astrophys. J.}, 495, L75

\item[]
Cappellaro, E., Turatto, M., Tsvetkov, D.Yu., Bartunov, O.S., Pollas, C.,
Evans, R., \& Namuy, M. 1997, {\it Astron. \& Astrophys.} 322, 431

\item[]
Chandrasekhar, S., 1961, {\it ``Hydrodynamic and hydromagnetic stability''}, 
Clarendon Press, Oxford, 1961

\item[] 
Ciotti, L., D'Ercole, A., Pellegrini, S., \& Renzini, A., 1991, 
{\it Astrophys. J.} 376, 380

\item[] 
Colella, P., \& Woodward, P.R., 1984, {\it J. Comput. Phys.} 54, 174

\item[]
Colella, P., \& Glaz, H.M., 1985, {\it J. Comput. Phys.} 59, 264

\item[]
Colless, M., \& Dunn, A.M., 1996, {\it Astrophys. J.} 458, 435

\item[]
De Grandi, S., \& Molendi, S., 2000, {\it Astrophys. J.}, in press,  
astro-ph/0012232

\item[]
Dehnen, W., \& Gerhard, O.E., 1994, {\it Monthly Not. of the Royal Astron. 
Soc.} 268, 1019

\item[]
Deiss, B.M., \& Just, A., 1996, {\it Astron. Astrophys.} 305, 407

\item[]
Dow, K.L., \& White, S.D.M., 1995,  {\it Astrophys. J.} 439, 113

\item[]
Drake, N., Merrifield, M.R., Sakelliou, I., \& Pinkney, J.C., 2000, 
{\it Monthly Not. of the Royal Astron. Soc.} 314, 768

\item[]
Dressler, A., \& Shectman, S.A., 1988, {\it Astronom. J.} 95, 985

\item[]
Fabbiano, G., Kim, D.-W., \& Trinchieri, G., 1992, {\it Astrophys. J. Suppl.} 
80, 531

\item[]
Finoguenov, A., David, L.P., \& Ponman, T.J., {\it Astronomische Nachrichten} 
320, 286

\item[]
Forman, W., Jones, C., \& Tucker, W., 1985, {\it Astrophys. J.} 293, 102

\item[]
Fryxell, B.A., M\"uller, E., \& Arnett, D., 1989, {\it MPA Report} 449

\item[] 
Fukazawa, Y., Makishima, K., Tamura, T., Ezawa, H., Xu, H., Ikebe, Y.,  
Kikuchi, K., \& Ohashi, T., 1998, {\it Publ. of the Astron. Soc. of Japan} 50, 187

\item[]
Gaetz, T.J., Salpeter, E.E., \& Shaviv, G., 1987, {\it Astrophys. J.} 316, 530

\item[]
Gerhard, O., Jeske, G., Saglia, R.P., \& Bender, R., 1998, {\it Monthly Not. 
of the Royal Astron. Soc.} 295, 197

\item[]
Godunov, S.K., Zabrodin, A.V., \& Prokopov, G.P., 1961, {\it U.S.S.R. Comput. 
Mathem. and Math. Physics} 1, 1187

\item[]
Goldstein, S. (editor), 1938, 
{\it Modern Developments in Fluid Dynamics}, Fluid Motion 
Panel of the Aeronautical Research Committee, Clarendon Press, Oxford, 1938

\item[]
Gunn, J.E., \& Gott, J.R., 1972, {\it Astrophys. J.} 176, 1


\item[]
Irwin, J.A., \& Bregman, J.N., 2000, astro-ph/0009273


\item[]
Irwin, J.A., \& Sarazin, C.L., 1998, {\it Astrophys. J.} 499, 650

\item[]
Jaffe, W., 1983, {\it Monthly Not. of the Royal Astron. Soc.} 202, 995

%
\item[]
Kritsuk, A.G., 1992, {\it Astron. \& Astrophys.} 261, 78

\item[]
Kritsuk, A.G., 1995, {\it Monthly Not. of the Royal Astron. Soc.}, 280, 319

\item[]
Kritsuk, A.G., B\"ohringer, H., \& M\"uller, E., 1998, {\it Monthly Not. of the 
Royal Astron. Soc.} 301, 343

\item[]
Lea, S.M., \& De Young, D.S., 1976, {\it Astrophys. J.} 210, 647

\item[]
Le Veque, R., ``{\it Nonlinear Conservation Laws and Finite Volume Methods 
for Astrophysical Fluid Flow}'', in {\it ``Computational methods for astrophysical 
fluid flow. Saas Fee Advanced Course 27''}, Lecture Notes 1997 of the Swiss 
Society for Astronomy and Astrophysics (SSAA), held March 3-8, 1997 in Les 
Diablerets, Switzerland. Edited by O. Steiner, and A. Gautschy. Publisher: Berlin, 
New York: Springer, 1998. 

\item[]
Loewenstein, M., \& Mathews, W.G., 1987, {\it Astrophys. J.} 319, 614

\item[]
Molendi, S., \& de Grandi, S., 1999, {\it Astron. \& Astrophys.} 351, L41

\item[]
Neumann, D.M., 1997, {\it ``R\"ontgenbeobachtungen von Galaxienhaufen und 
ihre Aussagen f\"ur die Kosmologie}, PhD Thesis at the Ludwig-Maximilians 
Universit\"at M\"unchen. 

\item[]
Nittmann, J., Falle, S.A.E.G., \& Gaskell, P.H., 1982, {\it Monthly Not. of 
the Royal Astron. Soc.} 201, 833

\item[]
Nulsen, P.E.J., 1982, {\it Monthly Not. of the Royal Astron. Soc.} 198, 1007

\item[]
Perry, A.E., \& Lim, T.T., 1978, {\it J. Fluid Mech.} 88, 451

\item[]
Portnoy, D., Pistinner, S., \& Shaviv, G., 1993, {\it Astrophys. J. Suppl.}  86, 
95 

\item[]
Rangarajan, F.V.N., White, D.A., Ebeling, H., \& Fabian, A.C., 1995, {\it Monthly 
Not. of the Royal Astron. Soc.} 277, 1047

\item[]
Reiprich, T.H., 1998, {\it ``Massenbestimmung an einer Stichprobe von 
Galaxienhaufen''}, Diplomarbeit, Ludwig-Maximilians-Universit\"at M\"unchen

\item[]
Renzini, A., 1988, in {\it Windows on Galaxies}, Eds. G. Fabbiano, J.S. 
Ghallagher \& A. Renzini, Dordrecht: Kluwer, p. 255

\item[]
Saglia, R.P., Bertin, G., \& Stiavelli, M., 1992, {\it Astrophys. J.} 384,433

\item[]
Sarazin, C.L., \& White, R.E., 1987, {\it Astrophys. J.} 320, 32

\item[]
Sarazin, C.L., \& Ashe, G.A., 1989, {\it Astrophys. J.} 345, 22

\item[]
Schindler, S., Binggeli, B., B\"ohringer, H., 1999, {\it Astron. \& 
Astrophys.} 343, 420

\item[]
Schipper, L., 1974, {\it Monthly Not. of the Royal Astron. Soc.} 168, 21

\item[]
Schlegel, E.M., Petre, R., \& Loewenstein, M., 1998, {\it Astronom. J.} 115, 525

%
\item[]
Sutherland, R.S., \& Dopita, M.A., 1993, {\it Astrophys. J. Suppl.} 88, 253 

\item[] 
Takeda H., Nulsen, P.E.J., \& Fabian, A.C., 1984, {\it Monthly Not. of the 
Royal Astron. Soc.} 208, 261

\item[]
Toniazzo, T., 1998, {\it ``Hydrodynamical simulation of the X-ray emitting gas
associated with cluster elliptical galaxies''}, PhD Thesis, Technische 
Universit\"at M\"unchen

\item[]
Trinchieri, G., Noris, L., \& di Serego Alighieri, S., 1997, {\it Astron. \& 
Astrophys.} 326, 565

\item[]
Wheeler, J.C., Sneden, C., \& Truran, J.W., 1989, {\it Ann. Rev. A. A.} 27, 279

\item[] 
White, R.E., \& Sarazin, C.L., 1991, {\it Astrophys. J.} 367, 476

\end{itemize}

\end{document}